\documentclass[twocolumn,preprintnumbers,groupedaddress,showpacs,showkeys]
{revtex4}
\usepackage{graphicx}

\def\to{\rightarrow}
\def\lr{\leftrightarrow}
\def\Re{{\rm Re}}
\def\lsim{\buildrel < \over {_\sim}}
     
\def\qb{{\bar{q}}}
\def\alphas{\alpha_s}
\def\eps{\epsilon}
\def\e{\epsilon}
\def\mgg{M_{\gamma\gamma}}
\def\pt{p_{\rm T}}
\def\et{E_{\rm T}}
\def\etmax{E_{\rm T\,max}}
\def\etjet{E_{\rm T\,jet}}
\def\Rjet{R_{\rm jet}}
\def\ptmiss{/{\hbox{\kern-6pt $\pt$}}}
\def\Ord{{\cal O}}
\def\MS{\rm\overline{MS}}
\def\Nf{{N_{\! f}}}
\def\cm{{\cal M}}
\def\Sublead{{\rm \scriptscriptstyle SL}}
\def\Lead{{\rm \scriptscriptstyle L}}
\def\spa#1.#2{\left\langle#1\,#2\right\rangle}
\def\spb#1.#2{\left[#1\,#2\right]}
\def\fig#1{fig.~{\ref{#1}}}
\def\Fig#1{Figure~{\ref{#1}}}
\def\eqn#1{eq.~(\ref{#1})}

\begin{document}

\preprint{hep-ph/0206194\\}
\preprint{UCLA/02/TEP/8}
\preprint{SLAC--PUB--9198}
\preprint{DAMTP-2002-41}
\preprint{MSUHEP--20522}

\title{Isolating a light Higgs boson \\
from the di-photon background at the LHC}

\author{Zvi Bern}
\thanks{Research supported by the US Department of Energy under grant
DE-FG03-91ER40662.}
\affiliation{Department of Physics and Astronomy \\
UCLA, Los Angeles, CA 90095-1547, USA}

\author{Lance Dixon}
\thanks{Research supported by the US Department of Energy under contract
DE-AC03-76SF00515.}
\affiliation{Stanford Linear Accelerator Center\\
Stanford University, Stanford, CA 94309, USA \\
and \\
Department of Applied Mathematics and Theoretical Physics\\
Wilberforce Road, Cambridge, CB3 0WA, UK}

\author{Carl Schmidt}
\thanks{Research supported by the US National Science Foundation under
grant PHY-0070443.}
\affiliation{Department of Physics and Astronomy \\
Michigan State University, East Lansing, MI 48824, USA}

\date{\today}

\begin{abstract}
We compute the QCD corrections to the gluon fusion subprocess $gg \to
\gamma\gamma$, which forms an important component of the background to
the search for a light Higgs boson at the LHC.  We study the
dependence of the improved $pp \to \gamma \gamma X$ background
calculation on the factorization and renormalization scales, on
various choices for photon isolation cuts, and on the rapidities of
the photons.  We also investigate ways to enhance the statistical
significance of the Higgs signal in the $\gamma\gamma$ channel.

\end{abstract}


\maketitle


\section{Introduction \label{IntroSection}}

The nature of electroweak symmetry breaking remains a mystery, despite
decades of theoretical and experimental study.  In the Standard Model,
the masses for the $W$ and $Z$ bosons, quarks and charged leptons are
all generated by the Higgs mechanism.  This mechanism leaves as its
residue the Higgs boson, the one undetected elementary particle of the
Standard Model, and the only scalar~\cite{Higgs}.  Its properties are
completely specified once its mass is determined.  Alternatives to, or
extensions of, the Standard Model electroweak symmetry breaking
mechanism typically also include one or more Higgs particles.
Experiments over the next decade at the Fermilab Tevatron and the CERN
Large Hadron Collider (LHC) should shed considerable light on
electroweak symmetry breaking, in particular by searching for these
Higgs bosons, and measuring their masses, production cross sections,
and branching ratios.

There are good reasons to believe that at least one Higgs particle
will be fairly light.  The Standard Model Higgs boson mass is bounded
from above by precision electroweak measurements, $m_H \lsim
196$--$230$ GeV at 95\% CL~\cite{HiggsRadCorr}.
In the Minimal Supersymmetric Standard Model (MSSM), the lightest Higgs
boson is predicted to have a mass below about 135 GeV~\cite{SusyHiggs};
over much of the parameter space it has properties reasonably similar to
the Standard Model Higgs boson.  There are also hints of a signal in the
direct search in $e^+e^- \to HZ$ at LEP2, just beyond the lower mass 
limit of 114.1~GeV~\cite{LEP2Limit}.  The corresponding 
lower limit on the lightest scalar in the MSSM is only 
91.0~GeV~\cite{LEP2MSSMLimit}, because the $HZZ$ coupling can be
suppressed in some regions of parameter space.

Run II of the Tevatron can exclude Standard Model Higgs masses up to
$\approx 180$~GeV with 15~fb$^{-1}$ per experiment.  However, at this
integrated luminosity a 5$\sigma$ discovery will be difficult to
obtain~\cite{RunIIExpectations} for a mass much beyond the LEP2 limit.
Also, the Higgs decay modes relevant for searches at the Tevatron, $H
\to b\bar{b}$ and $W^+W^-$, do not lend themselves to a precise
measurement of the Higgs mass.  The LHC will completely cover the low
mass region preferred by precision electroweak fits and the MSSM, as
well as much higher masses.  For $m_H < 140$ GeV, the most important
mode involves production via gluon fusion, $gg\to H$, followed by the
rare decay into two photons,
$H\to\gamma\gamma$~\cite{HggVertex,Higgsgammagamma}.  Although this
mode has a very large continuum $\gamma\gamma$
background~\cite{HBkgdgammagamma}, the narrow width of the Higgs
boson, combined with the mass resolution of order 1\% achievable in
the LHC detectors, allows one to measure the background experimentally
and subtract it from a putative signal
peak~\cite{ATLAS,CMS,Tisserand,Wielers}.

For the same mass range of $m_H < 140$ GeV at the LHC, Higgs
production via weak boson fusion, $qq' \to qq' W^+ W^- \to qq' H$,
followed by (virtual) weak boson decay, $H \to W^+ W^- \to e^\pm
\mu^\mp \ptmiss$, is also promising, even for Higgs masses as low as
the LEP2 limit~\cite{RZ}.  On the other hand, a mass determination
from this mode, or from weak boson fusion followed by $H \to
\tau^+\tau^-$~\cite{RZH}, cannot compete with the $\gamma\gamma$ mode,
although these modes certainly offer very useful branching ratio
information.

The purpose of this paper is to provide an improved calculation of the
prompt ({\it i.e.} not from hadron decay) di-photon background to
Higgs production at the LHC, in particular by computing QCD
corrections to the gluon fusion subprocess.  Although the background
will be measured at the LHC, it is still useful to have a robust
theoretical prediction in order to help validate the quantitative 
understanding of detector performance.  Perhaps more importantly,
the theoretical prediction can be used to systematically study the
dependence of the signal relative to the background on various kinematic
cuts, providing information which can be used to optimize Higgs search 
strategies.  As a side benefit, one can improve the predictions for a 
variety of di-photon distributions that can be measured.

\begin{figure*}
\includegraphics{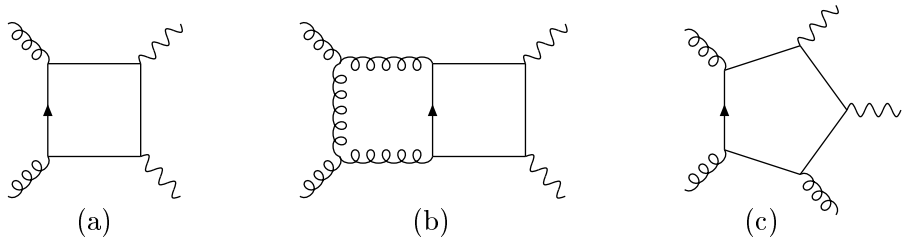}
\caption{\label{feyndiags} Sample quark loop diagrams contributing to 
$pp\to\gamma\gamma X$, which are computed in this paper:
(a) the leading order gluon fusion subprocess $gg \to \gamma\gamma$,
(b) the virtual correction to this subprocess,
and (c) the radiative process $gg \to \gamma\gamma g$.  }
\end{figure*}

The process $pp \to \gamma\gamma X$ proceeds at lowest order via the
quark annihilation subprocess $q\qb \to \gamma\gamma$, which is
independent of the strong coupling $\alphas$.  The
next-to-leading-order (NLO) corrections to this subprocess have been
incorporated into a number of Monte Carlo
programs~\cite{TwoPhotonBkgd1,DIPHOX}.  However, the gluon
distribution in the proton becomes very large at small $x$, making
formally higher order corrections involving gluon initial states very
significant for the production of low-mass systems at the LHC.  Gluons
can fuse to photon pairs through one-loop quark box diagrams such as
the one shown in \fig{feyndiags}(a).  The order $\alphas^2$
contribution to $pp \to \gamma\gamma X$ from $gg \to \gamma\gamma$ is
indeed comparable to the leading-order quark annihilation
contribution~\cite{HBkgdgammagamma,ADW,TwoPhotonBkgd1,DIPHOX}.  (In
fact, the NLO correction to quark annihilation, including the
$qg\to\gamma\gamma q$ amplitude, can also be as large as either of
these terms.)  Hence, to reduce the uncertainty on the total
$\gamma\gamma$ production rate, we have computed the contributions of
the $gg \to \gamma\gamma$ subprocess at {\it its}
next-to-leading-order, which we shall call ``NLO'', even though it is
formally N$^3$LO as far as the whole process $pp \to \gamma\gamma X$
is concerned.

The NLO gluon fusion computation has two matrix-element ingredients:
\begin{itemize}
\item The virtual corrections to $gg\to\gamma\gamma$,
involving two-loop diagrams such as the one in \fig{feyndiags}(b), 
which were computed recently~\cite{GGGamGam} using the integration 
methods developed in refs.~\cite{TwoloopIntegrals}.  
\item The effects of gluon bremsstrahlung, through the one-loop 
amplitude for $gg\to\gamma\gamma g$, including pentagon diagrams as 
depicted in \fig{feyndiags}(c).  The $gg\to\gamma\gamma g$ amplitude
can be obtained from the one-loop five-gluon matrix 
elements~\cite{FiveGluon} by summing over permutations of the 
external legs~\cite{GGGamGamGa,GGGamGamGb}.  
\end{itemize}
Both the virtual and real corrections have been evaluated in the limit
of vanishing quark masses.  In the range of di-photon invariant masses
relevant for the Standard Model and MSSM Higgs searches, 90--150 GeV,
this is an excellent approximation.  The masses of the five light
quarks are all much less than the scale of the process.  The top quark
contribution is negligible until the invariant mass approaches $2m_t
\approx 350$ GeV; at 150 GeV, it is still well under one percent of
the total quark loop contribution.  The virtual and real corrections
are separately infrared divergent.  We have used the dipole
formalism~\cite{CataniSeymour} to combine them into a finite result,
in a numerical program that can compute general kinematic
distributions.

%

Prompt photons are not only produced directly in hard processes, but
also {\it via} fragmentation from quarks and gluons.  As discussed in
ref.~\cite{DIPHOX}, even though the separation into direct and
fragmentation contributions is somewhat arbitrary, it is still very
useful to track the pieces separately.  The fragmentation processes
occur at low $\pt$ with respect to a neighboring jet.  They can only
be computed with the aid of nonperturbative information, in the form
of the quark and gluon fragmentation functions to a photon,
$D_{\gamma/(q\ {\rm or}\ g)}(z,\mu_F)$.  Here $z$ is the fractional
collinear momentum carried by the photon, measured relative to the
momentum of the parton which fragments into it, and $\mu_F$ is the
factorization scale used to separate the hard and soft processes.  The
{\it inclusive} di-photon production rate at the LHC is actually
dominated by the single fragmentation process, {\it e.g.}  the
partonic subprocess $qg\to\gamma q$, followed by the fragmentation
$q\to\gamma X$.  Even the double fragmentation process, {\it e.g.}
$gg \to q \qb$ followed by the fragmentations $q\to\gamma X$ and
$\qb\to\gamma X$, can exceed the direct contribution in the
80--140~GeV range for $\mgg$~\cite{DIPHOX}.

However, fragmentation contributions can be efficiently suppressed by
photon isolation cuts~\cite{DIPHOX}.  Such cuts are mandatory in order
to suppress the very large reducible experimental background where
photons are faked by jets, or more generally by hadrons.  In
particular, $\pi^0$s at large $\pt$ decay into two nearly collinear
photons, which can be difficult to distinguish from a single photon.
The standard method for defining an isolated photon is to first draw a
circle of radius $R$ in the plane of pseudorapidity, $\eta =
\ln\tan(\theta/2)$, and azimuthal angle, $\phi$, centered on the
photon candidate.  The amount of transverse hadronic energy in this
circle, or cone, is required to be less than some specified amount,
$\etmax$.  Here, $R$ and $\etmax$ may in principle be varied
independently.  Although photon isolation is improved by increasing
$R$ and decreasing $\etmax$, this cannot be done indefinitely, for
both theoretical and experimental reasons.  Theoretically, it is not
infrared-safe to forbid all gluon radiation in a finite patch of phase
space, so any prediction would have large uncontrolled corrections.
Experimentally, fluctuations in the number of soft hadrons from the
hard scattering, the underlying event, and other minimum bias events
in the same bunch crossing, plus detector noise, impose a lower limit
on the $\etmax$ that can be required for isolation, at a given $R$.

A typical choice of isolation criteria in past LHC studies has been
$R=0.4$ and $\etmax = 5$ or 15 GeV~\cite{DIPHOX}.  The experimental
optimization of these variables has often been made with the
suppression of the huge reducible background from jet fakes as a
primary criterion~\cite{ATLAS,CMS,Tisserand,Wielers}.  This criterion
is very understandable in the light of how poorly this background is
understood; it depends on the tails of distributions, such as the very
hard ($z \approx 1$) tail of parton fragmentation to
$\pi^0$s~\cite{PiBkgd}.  Nevertheless, it is estimated that this
background can be reduced to the order of 10--20\% of the irreducible
$\gamma\gamma$ background, so that one should try to optimize with
respect to the latter background as well.  We shall investigate the
behavior of the irreducible $\gamma\gamma$ background, as well as the
Higgs signal, as the cone size $R$ is increased, while also increasing
the transverse energy allowed into the cone.

An alternative ``smooth'' cone isolation criterion has been proposed by
Frixione~\cite{Frixione}, in which a continuous set of cones with $r<R$
are defined, and the transverse energy permitted inside $r$, $\et(r)$,
decreases to zero as $r$ does.  This criterion has the theoretical
advantage that it entirely suppresses the more poorly known fragmentation 
contribution, while still being infrared safe.  Experimentally, however, this
theoretical ideal may be difficult to achieve with a detector of finite
granularity, and taking into account the transverse extent of 
the photon's electromagnetic shower~\cite{Wielers}.  Nevertheless, we
shall study the behavior of Higgs signal and background for a few versions
of the smooth cone criterion as well.

The fragmentation contributions are technically involved to compute at
NLO.  Their implementation requires, for example, all the one-loop
four-parton matrix elements and tree-level five-parton matrix
elements, as well as convolution with fragmentation functions.
Fortunately, a flexible program is available, {\tt
DIPHOX}~\cite{DIPHOX,PiBkgd}, which incorporates at NLO the quark
annihilation direct subprocess, plus single and double fragmentation.
It also contains the gluon fusion subprocess at leading order.  We
have used {\tt DIPHOX} to produce all the non-gluon fusion
contributions to $pp\to\gamma\gamma X$ for the case of standard cone
isolation.  These were then combined with the results of our NLO
implementation of gluon fusion.  (We have also checked the {\tt
DIPHOX} implementation of the leading order gluon fusion contribution
against ours, and they agree perfectly.) For the case of smooth cone
isolation, where fragmentation contributions are absent, we have
implemented the $q\qb \to \gamma\gamma$ direct subprocess at NLO
ourselves, as well as the gluon fusion subprocess.


\begin{figure*}
\includegraphics{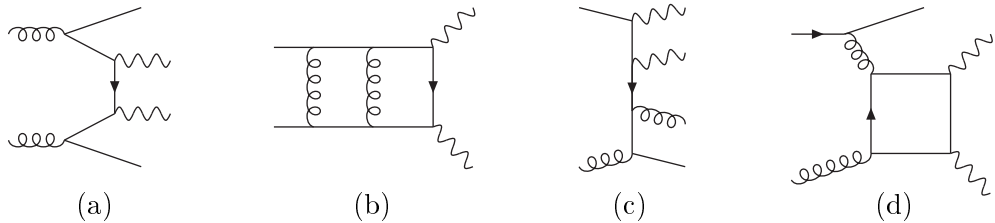}
\caption{\label{notincldiags} Sample diagrams for contributions
to $pp\to\gamma\gamma X$ which are {\it not} treated in this paper,
although they are of order $\alpha_s^2$:
(a) the tree-level subprocess $gg \to \gamma\gamma q\qb$,
(b) doubly virtual correction to $q\qb\to\gamma\gamma$,
(c) doubly real correction to $q\qb\to\gamma\gamma$, of the form
$qg\to \gamma\gamma gq$,
and (d) the process $qg \to \gamma\gamma q$ {\it via} a quark box.}
\end{figure*}
%


Since we are computing only a selection of the contributions to
$pp\to\gamma\gamma X$ at order $\alpha_s^2$ and $\alpha_s^3$, we
should make a few remarks about contributions we have omitted, at
least those appearing at order $\alpha_s^2$.  \Fig{notincldiags}
indicates a few such contributions.  Perhaps most worrisome at first
glance is the tree-level cross section for $gg \to \gamma\gamma q\qb$,
as shown in \fig{notincldiags}(a).  This subprocess can utilize the
large gluon-gluon luminosity, is only of order $\alpha_s^2$, and has
not yet been computed to our knowledge.  In addition, it can be
enhanced by $\gamma q$ collinear singularities if the isolation cut is
not too severe.  It has a double collinear singularity related to the
$g \to q\qb$ splitting process, so that this contribution should
depend on the factorization scale through terms proportional to
$\alpha_s^2(\ln{\mu_F})^n$ with $n=1,2$.  Although this contribution
alone does not cancel the full dependence on $\mu_F$ to order
$\alpha_s^2$, it does cancel the terms in $d\sigma/d(\ln\mu_F)$ that
are enhanced by two factors of the gluon density and are potentially
the largest.  Therefore, it is reasonable to consider the $gg \to
\gamma\gamma X$ contributions as a first approximation to the full
NNLO calculation.  A study of the similar subprocesses, $gg \to WV
q\qb'$ where $V=\gamma$ or $Z$, was recently carried out, with the
conclusion that their effects are actually quite small, 5\% or less
(and negative), with respect to the lower-order contributions of the
$q\qb$ and $qg$ initial states~\cite{AdFS}.  Assuming that these
results also hold for the $\gamma\gamma$ final state, our neglect of
$gg \to \gamma\gamma q\qb$ should be tolerable.

Other order $\alpha_s^2$ corrections to the quark annihilation
subprocess include the two-loop virtual corrections shown in
\fig{notincldiags}(b), which also have been computed
recently~\cite{QQGamGam}, and real corrections, such as from $qg\to
\gamma\gamma gq$ in \fig{notincldiags}(c).  The numerical
implementation of these corrections involve doubly unresolved parton
configurations and remain to be performed.  Because the
$qg\to\gamma\gamma q$ amplitude provides such a large NLO correction
to the quark annihilation subprocess, the order $\alphas^2$
contributions arising from the $qg$ initial state could be reasonably
important.  The $qg$ initial-state contributions may be more tractable
than the full order $\alpha_s^2$ corrections to the quark annihilation
subprocess, because of fewer soft gluon singularities.

There are actually three other ways for the quark box responsible for
$gg\to\gamma\gamma$ to contribute to the $pp\to\gamma\gamma X$ cross
section at order $\alphas^2$.  The quark box can be incorporated into
the one-loop $qg \to \gamma\gamma q$ amplitude shown in
\fig{notincldiags}(d), which can then interfere with the tree-level
$qg \to \gamma\gamma q$ amplitude.  However, such a contribution only
gets to utilize one gluon distribution function at small $x$, not two.
It is likely to be smaller than the other order $\alpha_s^2$ $qg$
initial-state contributions just mentioned, given that it lacks both
initial- and final-state singularities.  (The loop and tree amplitudes
each have square-root collinear singularities, but they appear in
different collinear limits, so that the phase space integral of the
amplitude product remains finite.)  The quark box contribution to the
crossed process $q\qb\to\gamma\gamma g$ is also infrared finite, and
lacks any small $x$ gluon enhancement, so it should be even smaller
than the $qg \to \gamma\gamma q$ case.  Finally, there is a two-loop
virtual correction to $q\qb \to \gamma\gamma$ containing the quark box
(not shown here), which can interfere with the tree amplitude at order
$\alphas^2$.  This particular correction is also infrared finite, and
does not benefit from any small $x$ gluon distribution.  Numerical
evaluation of the expression in ref.~\cite{QQGamGam} shows that its
magnitude never exceeds 0.3\% of the Born quark annihilation cross
section (its sign is negative for all relevant, central scattering
angles), so we are justified in neglecting it.

In this paper we also investigate the effects of isolation cuts and
other kinematic features of the Higgs signal {\it vs.} the QCD
background. Intuitively, one might expect the irreducible background
to diminish more rapidly than the signal as one imposes more severe
photon isolation requirements for the following reason: The largest
component of the irreducible background, for typical isolation cuts,
comes from the subprocess $qg \to \gamma\gamma q$, even after we
include the NLO corrections to gluon fusion.  The $qg$-initiated
subprocess has final-state singularities when the quark is collinear
with either of the photons, which of course require some kind of
photon isolation to make finite. Thus, the contribution of this
subprocess can be significantly reduced with tighter isolation cuts.
On the other hand,
the dominant Higgs production process, $gg \to H X$, even at higher
orders where additional partons are radiated, clearly does not give
rise to partons that are preferentially near the photons.

\begin{figure}
\includegraphics[width=3.cm]{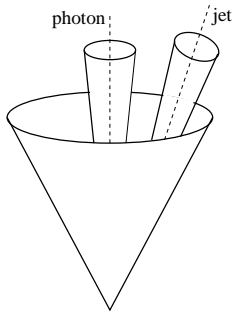}
\caption{\label{figjetveto} 
Illustration of two possible ways to strengthen photon isolation, beyond
the ``standard'' isolation represented by the inner cone.
One can either increase the cone radius to the large outer one shown,
or one can veto on jets within such a radius.}
\end{figure}

There are at least two different ways to make the photon isolation
more severe.  One way is to strengthen the isolation cone, for example
by increasing its radius $R$.  A second approach is to impose a jet
veto~\cite{Tisserand} in the neighborhood of the photon candidate, on
top of a ``standard'' isolation requirement.  For example, one may
forbid events having a jet with $\et > \etjet$ whose axis is within a
distance $\Rjet$ in $(\eta,\phi)$ space from either photon candidate,
where $\Rjet > R$.  The two approaches are depicted in
\fig{figjetveto}.  The small inner cone corresponds to a standard
photon isolation cone, {\it e.g.} with radius $R=0.4$.  The outer cone
could correspond to a larger isolation cone, {\it e.g.} with radius
$R=1$ or $R=2$.  Alternatively, it could represent the radius $\Rjet$
within which jets, such as the one depicted, are vetoed against.

At the level of a NLO parton calculation, the jet veto approach is not
too different from the large cone approach, if one chooses $\Rjet$ and
$\etjet$ to be equal to the large cone values of $R$ and $\etmax$,
respectively.  The reason is that for the direct contributions the
hadronic energy is deposited as a single parton.  However, the jet
veto approach is probably preferable for experimental reasons.  The
hadronic energy represented by the final state quark in the background
process $qg \to \gamma\gamma q$ should typically be deposited in the
form of a jet.  In contrast, an equivalent amount of energy
originating from either soft hadrons from overlapping events, or
detector noise, will usually be uniformly distributed in
$(\eta,\phi)$, and so will be less likely to form a jet.  Thus when a
signal event is accompanied by such energy, it is less likely to be
discarded in the case of a jet veto, compared with the case of a
standard or smooth isolation cone of large size.

Another feature distinguishing the two photons from Higgs decay from
those produced directly is their angular distribution.  The $q\qb$-
and $qg$-initiated components of the background have $t$-channel
fermion exchanges that tend to yield photons toward smaller scattering
angles in the $\gamma\gamma$ center-of-mass (for a fixed $\mgg$),
compared with the decay of the Higgs boson, which is isotropic in
$\cos\theta^*$.  Hence, we will study signal and background
distributions in $y^* \equiv (y(\gamma_1) - y(\gamma_2))/2$, where the
$y(\gamma_i)$ are the rapidities of the photons.  This variable serves
as a proxy for $\cos\theta^*$, because $\cos\theta^* = \tanh y^*$ at
lowest order.

The remaining variable which describes the Higgs boson kinematics at
LO is the rapidity of the Higgs boson itself.  This leads us to
consider the signal and background distributions of $y_{\gamma
\gamma}$, the rapidity of the total di-photon system.  To a large
extent this distribution is determined by the parton luminosities
which are involved in the production process.  Since Higgs production
is dominated by gluon-gluon fusion, whereas the direct $\gamma \gamma$
background gets comparable contributions from all three initial
states, $q\bar q$, $qg$, and $gg$, there may be a useful difference in
the $y_{\gamma \gamma}$ distribution.

To carry out these studies of the signal as well, we have implemented
the gluon fusion production of the Standard Model Higgs boson at NLO,
followed by decay to $\gamma\gamma$.  We work in the heavy top quark
limit, for which an effective $Hgg$ vertex~\cite{HggVertex} suffices
to describe the production process at the low Higgs transverse momenta
we consider.  (We do, however, include the exact $m_H/m_t$ dependence
of the LO term. This has been shown~\cite{NLOHKfactor} to be an
excellent approximation to the exact NLO cross section~\cite{NLOHiggs}
for the Higgs masses with which we are concerned.)  The total Higgs
production cross section has recently become available in this limit
at NNLO~\cite{HKNNLO}.  The corrections from NLO to NNLO are modest.
Here we wish to study distributions with kinematical cuts, which have
not yet been computed at NNLO.  For cuts that select events with
nonzero transverse momentum of the Higgs boson (which we do not
explicitly consider in this paper), our calculation is only a LO
computation; distributions for such quantities at NLO are available
elsewhere~\cite{dFGKRSVN}.  For the branching ratio ${\rm Br}(H \to
\gamma\gamma)$ we use the program {\tt HDECAY}~\cite{HDECAY}.

This paper is organized as follows.  In
section~\ref{ComputationSection} we outline the NLO gluon fusion
computation.  In section~\ref{NumericsSection} we show the effects of
the NLO corrections to $gg\to\gamma\gamma$ on the total
$pp\to\gamma\gamma X$ cross section, using kinematic cuts appropriate
to the Higgs search, and concentrating on the comparison with LO and
on the scale dependence.  In section~\ref{StatisticsSection} we study
the effects of varying the photon isolation criteria, either by
increasing the size of the isolation cone, or by imposing a veto on
nearby jets.  We also compute the distributions in $y^*$ and
$y_{\gamma \gamma}$, and investigate how they may be used to help
distinguish the Higgs signal from the background.  In
section~\ref{ConclusionSection} we present our conclusions, and the
outlook for further improvement of our understanding of the
$\gamma\gamma$ background.


\section{Outline of the Computation \label{ComputationSection}}

The gluon fusion subprocess begins at one loop.  The next-to-leading
order corrections to it include two-loop four-point amplitudes and
one-loop five-point amplitudes.  The calculation of these matrix
elements requires a fair amount of
work~\cite{GGGamGam,FiveGluon,GGGamGamGa,GGGamGamGb}.  However, once
the amplitudes are available, the rest of the NLO gluon fusion
computation is in all essential respects identical to a conventional
NLO computation involving tree-level and one-loop amplitudes.  For
example, the two-loop amplitudes for $gg\to\gamma\gamma$ have virtual
infrared divergences with the same structure as those of a typical
one-loop amplitude.  Similarly, the vanishing of the
$gg\to\gamma\gamma$ tree amplitude causes the singular behavior of the
$gg\to\gamma\gamma g$ one-loop amplitude, as the outgoing gluon
becomes soft or collinear with an incoming gluon, to have the same
form as that of a typical tree amplitude with one additional radiated
gluon.

The one-loop gluon fusion helicity amplitudes for the process
\begin{equation}
 g(-p_1,-\lambda_1) + g(-p_2,-\lambda_2)
\to \gamma(p_3,\lambda_3) + \gamma(p_4,\lambda_4), 
\label{gggamgamlabel}
\end{equation}
in an ``all-outgoing'' labelling convention for the momenta $p_i$ and
helicities $\lambda_i$, are given by 
\begin{equation}
\cm_{gg \to \gamma \gamma}^{1-{\rm loop}} =
4 \alpha \alphas(\mu_R) 
\, \delta^{a_1a_2} \biggl(\sum_{j = 1}^\Nf Q_j^2 \biggr)
M^{(1)}_{\lambda_1\lambda_2\lambda_3\lambda_4} \,,
\label{LOdecomp}
\end{equation}
where $\alpha$ is the QED coupling at zero momentum transfer, $\alpha
= 1/137.036\ldots$ and $\alphas(\mu_R)$ is the running QCD coupling in
$\MS$ scheme, evaluated at renormalization scale $\mu_R$.  Also, in
this formula, $a_{1,2}$ are the adjoint gluon color indices, $Q_j$ are
the quark charges in units of $e$, the appropriate number of light
flavors is $\Nf=5$, and we have suppressed overall phases in the
amplitudes.  The quantities
$M^{(1)}_{\lambda_1\lambda_2\lambda_3\lambda_4}$ are~\cite{GGGamGam}
\begin{eqnarray}
M_{++++}^{(1)} &=&
   1
\,, \nonumber \\
M_{-+++}^{(1)} &=& M_{+-++}^{(1)} = M_{++-+}^{(1)} = M_{+++-}^{(1)} =
   1
\,,  \nonumber \\
M_{--++}^{(1)} &=&
- {1\over2} {t^2+u^2\over s^2}
                  \Bigl[ \ln^2\Bigl({t\over u}\Bigr) + \pi^2 \Bigr]
\nonumber \\ && \hskip2pt
		- {t-u\over s} \ln\Bigl({t\over u}\Bigr) - 1
\,,  \nonumber \\
M_{-+-+}^{(1)} &=&
 - {1\over2} {t^2+s^2\over u^2}
                  \ln^2\Bigl(-{t\over s}\Bigr)
\nonumber \\ && \hskip2pt
		- {t-s\over u} \ln\Bigl(-{t\over s}\Bigr) - 1 \nonumber \\
&& \hskip2pt
	- i \pi \biggl[ {t^2+s^2\over u^2} \ln\Bigl(-{t\over s}\Bigr)
                      + {t-s\over u} \biggr]
\,, \nonumber \\
M_{+--+}^{(1)} &=&
M_{-+-+}^{(1)} \big|_{t\lr u}
\,,
\label{OneLoopFunctions}
\end{eqnarray}
where $s=(p_1+p_2)^2$, $t=(p_2+p_3)^2$, $u=(p_1+p_3)^2$, and the remaining
helicity amplitudes are obtained by parity.

The dimensionally-regularized and renormalized two-loop QCD corrections 
to these amplitudes can be written as~\cite{GGGamGam}
\begin{eqnarray}
\cm_{gg \to \gamma \gamma}^{2-{\rm loop}}
 &=&  { 2 \alpha \alphas^2(\mu_R) \over \pi } \delta^{a_1a_2} 
 \biggl(\sum_{j = 1}^\Nf Q_j^2 \biggr)
\nonumber \\ && \hskip-50pt  
\times \biggl\{  
\Bigl[ I^{(1)}(\e) 
    + b_0 \Bigl( \ln\Bigl({\mu_R^2\over s}\Bigr) + i\pi \Bigr) 
 \Bigr] M^{(1)}_{\lambda_1\lambda_2\lambda_3\lambda_4}
\nonumber \\ && \hskip-35pt
+ N F^\Lead_{\lambda_1\lambda_2\lambda_3\lambda_4}(s,t) 
- {1\over N} F^\Sublead_{\lambda_1\lambda_2\lambda_3\lambda_4}(s,t)
\biggr\} \,,
\label{TwoLoopDecomp}
\end{eqnarray}
where the number of dimensions is $D=4-2\e$, and all the poles in 
$\e$ are contained in 
\begin{equation}
I^{(1)}(\e) = 
- {e^{-\e \psi(1)} \over \Gamma(1-\e)} 
\Bigl[ {N \over \e^2} 
     + {b_0 \over \e}  \Bigr]
\Bigl({\mu_R^2\over -s} \Bigr)^{\e} \,,
\label{I0def}
\end{equation}
with
\begin{equation}
b_0 = { 11 N - 2 \Nf \over 6}\,, 
\label{b0def}
\end{equation}
and the number of colors is $N=3$.
The finite expressions 
$F^\Lead_{\lambda_1\lambda_2\lambda_3\lambda_4}(s,t)$
and $F^\Sublead_{\lambda_1\lambda_2\lambda_3\lambda_4}(s,t)$ are 
presented in ref.~\cite{GGGamGam}.

The radiative process,
\begin{eqnarray}
 && g(-p_1,-\lambda_1) + g(-p_2,-\lambda_2)
\nonumber \\ && \hskip20pt
 \to \gamma(p_3,\lambda_3) + \gamma(p_4,\lambda_4) + g(p_5,\lambda_5),
\label{gggamgamglabel}
\end{eqnarray}
begins at order $\alphas^3$.  The squared matrix element,
averaged over initial helicities and colors, and summed over final ones, 
and with a 1/2 for identical final-state photons, is given 
by~\cite{GGGamGamGa,GGGamGamGb}
\begin{eqnarray}
&&\overline{ | \cm |^2 }_{\rm \hskip-4pt rad} \equiv
\sum_{\rm hel., color}^{\rule{5mm}{.2mm}}
| \cm_{gg \to \gamma \gamma g}^{1-{\rm loop}} |^2
\nonumber \\
&& =\ 4 \pi \alpha^2 \alphas^3(\mu_R) { N \over N^2-1 }
  \biggl(\sum_{j = 1}^\Nf Q_j^2 \biggr)^2 
\nonumber \\ && \hskip10pt
\times \sum_{\rm hel.}
  \Biggl| \sum_{\sigma\in {\rm COP}_4^{(125)}}
  A_{5;1}^{[1/2]}(\sigma_1,\sigma_2,\sigma_3,\sigma_4,\sigma_5) \Biggr|^2 
 \,,
\label{gggamgamgsum}
\end{eqnarray}
where $\sigma_i$ label the helicities and momenta of the gluons and
photons, and ${\rm COP}_4^{(125)}$ denotes the subset of 12 permutations
of (1,2,3,4,5) that leave 5 fixed and preserve the cyclic ordering of
(1,2,5), {\it i.e.} 1 appears before 2.
The partial amplitudes
$A_{5;1}^{[1/2]}(\sigma)$ are those for five-gluon scattering {\it via} 
a quark loop given in ref.~\cite{FiveGluon}, but with an overall factor
of $(4\pi)^{-2}$ removed.  

The permutation sum in \eqn{gggamgamgsum} cancels out all
of the virtual divergences of the partial amplitudes, and most of their 
singularities as momenta become soft and collinear.  The only remaining
singularities are when the final gluon momentum $p_5$ becomes soft,
or becomes collinear with either initial gluon momentum, $p_1$ or $p_2$.
In the region, for example, where $p_5$ is collinear with $p_1$, with
$p_5 \to -(1-x) p_1$, the squared matrix element has the limiting
behavior,
\begin{equation}
\overline{ | \cm |^2 }_{\rm \hskip-4pt rad}\ \rightarrow\
\overline{ | \cm |^2 }_{\rm \hskip-4pt dipole,1},
\label{collinearbehavior}
\end{equation}
where
\begin{eqnarray}
&& \hskip-30pt
\overline{ | \cm |^2 }_{\rm \hskip-4pt dipole,1}
\equiv
  { - 1 \over 2 x p_1 \cdot p_5 } \biggl\{ P_{gg}(x)
  \times \sum_{\rm hel., color}^{\rule{5mm}{.2mm}}
| \cm_{gg \to \gamma \gamma}^{1-{\rm loop}} |^2 
\nonumber \\ && \hskip-30pt
 - 2N { 1-x \over x } \Re \biggl[ 
         { \spa1.5 \spb5.2 \spa2.1 \over \spb1.5 \spa5.2 \spb2.1 }
     \sum_{\rm hel., color}^{\rule{5mm}{.2mm} '} 
| \cm_{gg \to \gamma \gamma}^{1-{\rm loop}} |^2 \biggr] \biggr\} \,, 
\label{dipole1}
\end{eqnarray}
with
\begin{equation}
  P_{gg}(x) = 2 N \biggl[ {x \over 1-x } + { 1-x \over x} + x(1-x) \biggr]
  \,,
\label{Pgg}
\end{equation}
and the $gg\to \gamma\gamma$ process has the kinematics 
$(-xp_1) + (-p_2) \to p_3+p_4$.  The second term involves
the spinor products $\spa{i}.j$~\cite{MPReview}
entering the five-point partial amplitudes. 
It accounts for nontrivial phase behavior of the amplitudes 
as $p_5$ rotates azimuthally about the $p_1$ direction. The primed sum is 
defined analogously to the unprimed sum, except
that in each complex conjugated helicity amplitude 
the helicity of the gluon with momentum $(-xp_1)$ 
is flipped, so that 
$M^{(1)}_{\lambda_1\lambda_2\lambda_3\lambda_4}
 M^{(1) \, *}_{(-\lambda_1)\lambda_2\lambda_3\lambda_4}$ appears,
and the spinor product overall phases in the amplitudes should also be
included~\cite{GGGamGam}.

Integration over these singular phase space 
regions can be handled by the dipole formalism~\cite{CataniSeymour},
with just two dipole subtractions, one for each initial gluon.
The dipole subtraction for gluon 1 is given by \eqn{dipole1},
where $x = (p_1+p_2+p_5)^2/(p_1+p_2)^2$, and the four-point
matrix-elements are evaluated for boosted kinematics: gluon 1 is assigned
momentum $(-xp_1)$, gluon 2 momentum $-p_2$ still, and the photon momenta
$p_j$, $j=3,4$, are set equal to 
\begin{equation}
 \tilde{p}_j^\mu = p_j^\mu 
  - { 2p_j \cdot (K+\tilde{K}) \over (K+\tilde{K})^2 }
              (K+\tilde{K})^\mu
  + { 2p_j \cdot K \over K^2 } \tilde{K}^\mu \,,
\label{CSBoost}
\end{equation}
where 
\begin{eqnarray}
 K^\mu &=& - p_1^\mu - p_2^\mu - p_5^\mu \,,
\nonumber \\
 \tilde{K}^\mu &=& - x p_1^\mu - p_2^\mu \,.
\label{CSKDef}
\end{eqnarray}
Subtracting these dipole terms from the $gg \to \gamma\gamma g$
squared matrix element, \eqn{gggamgamgsum}, removes the soft and
collinear singularities, so that the resulting expression
can be integrated directly in four dimensions.
Furthermore, the dipole subtraction terms themselves can be 
integrated analytically in $D=4-2\epsilon$ dimensions~\cite{CataniSeymour}; 
their poles in $\e$, combined with the collinear counterterm in the 
$\MS$ factorization scheme, cancel against the virtual
divergence in the interference of 
$\cm_{gg \to \gamma \gamma}^{2-{\rm loop}}$ in \eqn{TwoLoopDecomp}
with $\cm_{gg \to \gamma \gamma}^{1-{\rm loop}}$.
Thus, by adding and subtracting the dipole terms, one obtains
an expression for the NLO cross section which is explicitly finite in 
four dimensions.

The remaining finite NLO differential 
cross section for $pp \to gg \to \gamma \gamma X$
consists of three terms, which can be put in the form
\begin{eqnarray}
 {d\sigma^{\rm NLO}\over dM^2_{\gamma\gamma}} &=& \int {dx_1 
dx_2dz\over\hat s} \, g(x_1) g(x_2) \delta\left(z-{M^2_{\gamma\gamma}/ 
\hat s}\right)\nonumber\\
&&\times\Bigl[\delta(1-z)d\hat\sigma^{\rm B}+
d\hat\sigma^{\rm C}+d\hat\sigma^{\rm R}\Bigr],
\label{diffcross}
\end{eqnarray}
where the three terms are functions of the incoming parton 
momenta $x_1p_a$ and $x_2p_b$ and the variable $z$, and we have used
$\hat s=(x_1p_a+x_2p_b)^2$.
The term with leading-order kinematics~(\ref{gggamgamlabel}), 
including the LO answer, can be written
\begin{eqnarray}
d\hat\sigma^{\rm B} &=&
    d\hat\sigma^{\rm LO}
\biggl( 1 + { \alphas(\mu_R) \over \pi } 
      \Bigl[ 2 b_0 \ln\Bigl( {\mu_R \over \mu_F} \Bigr)
             + { \pi^2 \over 3} N \Bigr] \biggr)
\nonumber \\ && \hskip-35pt
 +{ 1 \over 2 
\hat s  } 2 \, \Re \biggl[ \sum_{\rm hel., color}^{\rule{5mm}{.2mm}}
 \cm_{gg \to \gamma \gamma}^{2-{\rm loop,\,fin.}}
   \cm_{gg \to \gamma \gamma}^{1-{\rm loop} \, *} \biggr]d\Gamma_2 \,,
\label{Bornterm}
\end{eqnarray}
where     
\begin{equation}
d\hat\sigma^{\rm LO}\equiv d\hat\sigma^{\rm LO}(x_1p_a,x_2p_b) = 
 { 1 \over 2\hat s }
 \sum_{\rm hel., color}^{\rule{5mm}{.2mm}}
 | \cm_{gg \to \gamma \gamma}^{1-{\rm loop}} |^2 d\Gamma_2\,,
\label{LOSigmaDef}
\end{equation}
$d\Gamma_2$ is the two-particle Lorentz-invariant phase space,
and $\cm_{gg \to \gamma \gamma}^{2-{\rm loop,\,fin.}}$ refers to 
just the terms containing $F^\Lead$ and $F^\Sublead$
in~\eqn{TwoLoopDecomp}.
The second term also has leading-order kinematics, but boosted along the 
beam axis, with
\begin{eqnarray}
d\hat\sigma^{\rm C}&=& 
   { \alphas(\mu_R) \over 2\pi } z\tilde{K}^{g,g}(z) \nonumber\\
&&\hskip-30pt 
\times\Bigl[ d\hat\sigma^{\rm LO}(zx_1p_a,x_2p_b)+ 
 d\hat\sigma^{\rm LO}(x_1p_a,zx_2p_b) \bigr]\,,
\label{Collinearconvolute}
\end{eqnarray}
where
\begin{eqnarray}
\tilde{K}^{g,g}(z) &=& 2 N \Biggl[  
2 \biggl( { \ln\bigl( (1-z) \mgg/ \mu_F \bigr) \over 1 - z } \biggr)_{+}
- { \ln z \over 1-z }
\nonumber \\ &&\hskip-40pt
+ \biggl[ {1-z \over z} - 1 + z(1-z) \biggr] 
   \ln\biggl( { (1-z)^2 \over z} { \mgg^2 \over \mu_F^2 } \biggr) \Biggr]\,.
\label{tildeKdef}
\end{eqnarray}
The third term contains the $gg \to \gamma\gamma g$ squared matrix 
element~(\ref{gggamgamgsum}), minus the two dipole subtractions 
mentioned above,
\begin{equation}
 d \hat\sigma^{\rm R} =
 { 1 \over 2\hat s} \biggl[
   \overline{ | \cm |^2 }_{\rm \hskip-4pt rad}
 - \overline{ | \cm |^2 }_{\rm \hskip-4pt dipole,1}
 - \overline{ | \cm |^2 }_{\rm \hskip-4pt dipole,2} \biggr]d\Gamma_3,
\label{dipolesubtracted}
\end{equation}
where $d\Gamma_3$ is the three-particle Lorentz-invariant phase space.
Thus it involves the full five-point kinematics.

We have implemented these three terms in a numerical program that
allows for different kinematic cuts and photon isolation criteria to
be applied.  The numerical integrals are performed by adaptive Monte
Carlo sampling using {\tt VEGAS}~\cite{VEGAS}.  For the first and
second terms the Monte Carlo routine is used to generate the photon
four-momenta with four-point kinematics, along with the appropriate
longitudinal boosts of \eqn{Collinearconvolute}.  For the third term
the events are generated with five-point kinematics; however,
treatment of the final-state particles differs for the $gg \to
\gamma\gamma g$ squared matrix element~\eqn{collinearbehavior} and the
two dipole subtractions.  Whereas the $gg \to \gamma\gamma g$ squared
matrix element is treated as a true three-parton final-state, the
dipole subtraction for gluon 1 is treated as a two-parton final-state
with the photon momenta given by \eqn{CSBoost}, and similarly for the
dipole subtraction for gluon 2.  Thus, each call to {\tt VEGAS} in the
third term produces three distinct kinematic configurations with three
different weights.  The infrared safety of the
kinematic and isolation cuts ensures the appropriate cancellation
between the pieces as the gluon becomes soft or collinear.


\section{Results for Di-photon Background  \label{NumericsSection}}

In this section we present the modifications to the $p p \rightarrow
\gamma \gamma X$ cross section due to the inclusion of the NLO
contributions to the gluon fusion subprocess.

\subsection{General remarks \label{NumericsGeneralSubsection}}

We impose the following kinematic cuts on the two photons,
\begin{equation}
\pt(\gamma_1) > 40\hbox{ GeV}, \quad \pt(\gamma_2) > 25\hbox{ GeV},
\quad | y(\gamma_{1,2}) | < 2.5,
\label{AccCuts}
\end{equation}
which are essentially those used by the ATLAS and CMS detectors in
their Higgs search studies~\cite{ATLAS,CMS}.  In addition, we require
each photon to be isolated from hadronic energy, according to one of
two criteria:
\begin{enumerate}
\item {\it standard} cone isolation --- the amount of transverse hadronic
energy $\et$ in a cone of radius 
$R = \sqrt{(\Delta\eta)^2 + (\Delta\phi)^2}$ must be less than $\etmax$.
\item {\it smooth} cone isolation~\cite{Frixione} 
--- the amount of transverse hadronic energy $\et$ in {\it all} cones 
of radius $r$ with $r < R$ must be less than 
\begin{equation}
\etmax(r) \equiv 
\pt(\gamma) \, \eps \, \biggl( { 1 - \cos r \over 1 - \cos R } \biggl)^n.
\end{equation}
Here we shall choose $n=1$.
\end{enumerate}
Note that in an NLO calculation, ``hadronic energy'' in the neighborhood 
of the photon always amounts to just a single parton, a rather crude
approximation to the true hadronic background.

Unless otherwise specified, we evaluate the NLO cross sections with
the MRST99 set 2 parton distributions~\cite{MRST99}, with the
corresponding value of $\alphas(m_Z) = 0.1175$.  For comparison
purposes, we also present cross sections for the leading-order gluon
fusion subprocess, convoluted with the NLO parton distributions MRST99
set 2.  We use NLO instead of LO parton distributions here, because
that approach was taken in the most complete previous
study~\cite{DIPHOX}, and we wish to highlight differences with respect
to that work.  The use of NLO parton distributions can also be
justified by considering the ``LO'' gluon fusion subprocess as a NNLO
correction to the entire $\gamma\gamma$ production process.

Our default choices for the renormalization and factorization scales
are
\begin{equation}
\mu_R = \mu_F = 0.5 \mgg,
\label{defaultmu}
\end{equation}
as in ref.~\cite{DIPHOX}; we also investigate the dependence of the
results on $\mu_R$ and $\mu_F$.
For the fragmentation contributions~\cite{DIPHOX}, we rely on the NLO set I
photon fragmentation functions from ref.~\cite{NLOFrag}.


\subsection{Effects of NLO gluon fusion on the $\gamma\gamma$ 
background \label{TotalCrossSubsection}}

\begin{figure*}
\includegraphics[width=8.65cm]{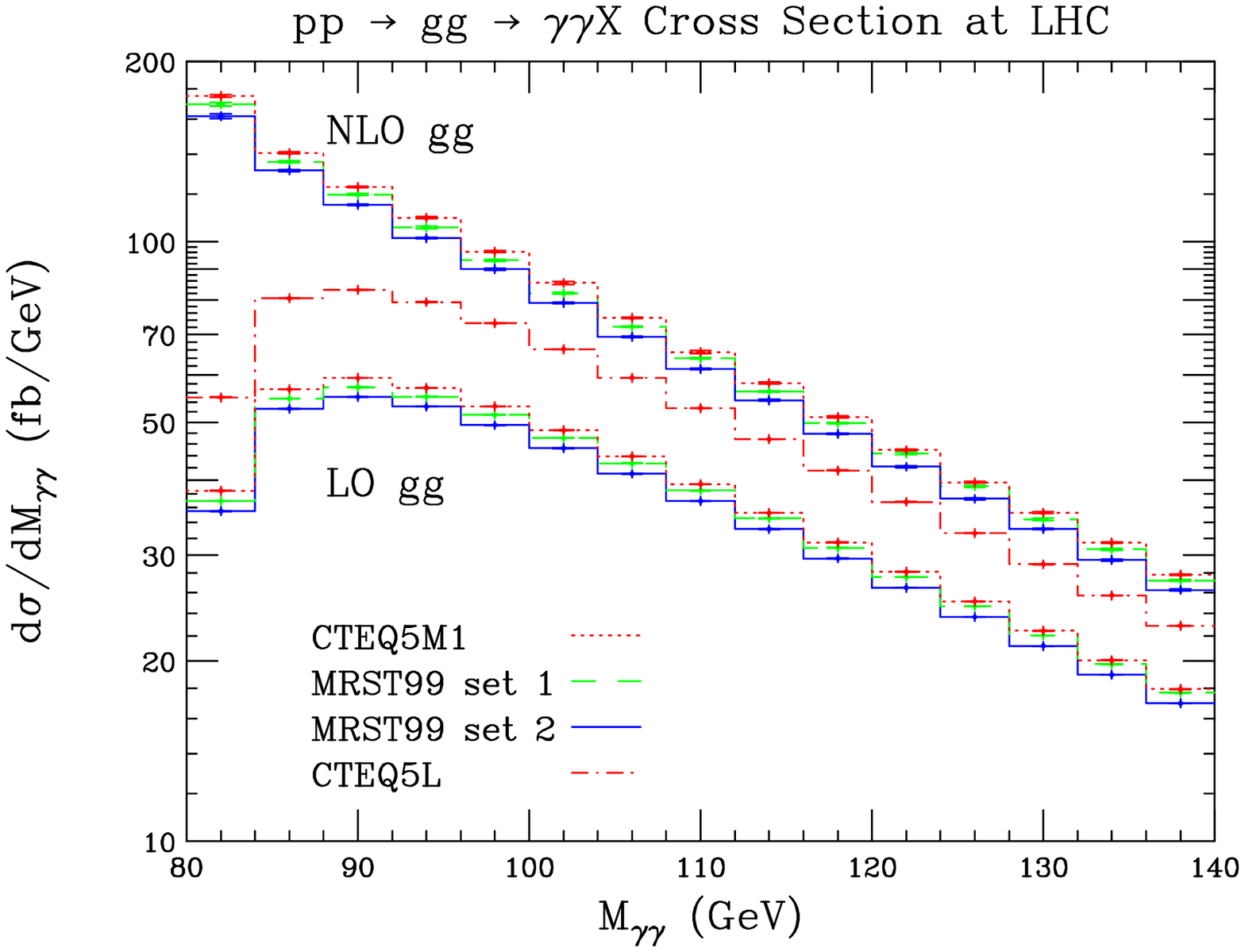}
\includegraphics[width=8.65cm]{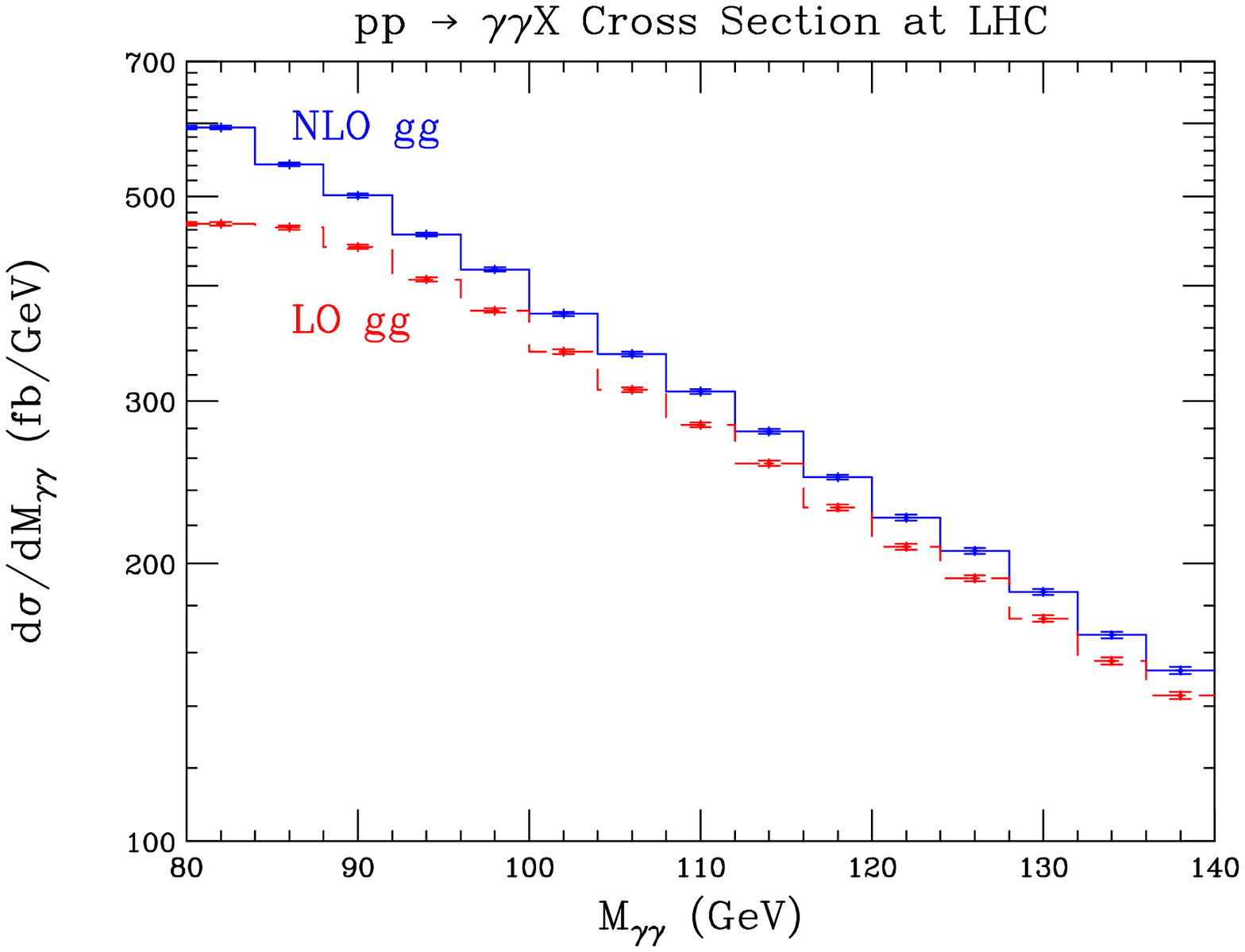}
\caption{\label{fig12_rp4} (a) Contribution of the gluon fusion
subprocess to $pp\to\gamma\gamma X$ at the LHC, at leading order
(lower four), and at NLO (upper three), using various parton
distributions.  (b) Total $pp \to \gamma\gamma X$ production at NLO,
including NLO $q\qb\to\gamma\gamma$ and fragmentation contributions,
with the gluon fusion subprocess treated at LO (dashed) and at NLO
(solid).  MRST99 set 2 partons are used in (b). Contributions not
involving gluon fusion into photons are obtained from {\tt DIPHOX}.
Both sets of plots are for $\mu_R = \mu_F = 0.5 \mgg$, and a standard
photon isolation criterion with $R=0.4$, $\etmax = 15$~GeV.
Statistical errors from numerical integration are shown.}
\end{figure*}

\Fig{fig12_rp4}(a) shows the contribution of just the gluon fusion
subprocess to $pp\to\gamma\gamma X$ at the LHC, at its leading and
next-to-leading orders, for the standard cone photon isolation
criterion with $R=0.4$, $\etmax = 15$~GeV, and for several choices of
parton distributions.  To help with comparisons to the results of
ref.~\cite{DIPHOX}, we use MRST99 set 2 as our ``default''
choice. This set has a somewhat larger gluon distribution at large $x$
than MRST99 set 1, but the differences with this set, or with
CTEQ5M1~\cite{CTEQ}, at the smaller $x$ ranges probed here are small
compared to the NLO corrections, or to the renormalization and
factorization scale dependence, as we shall see.  We also plot the LO
cross section with the LO CTEQ5L distributions (using a LO
$\alphas(m_Z)=0.127$)~\cite{CTEQ}, in order to compare our NLO cross
section with a ``true'' LO calculation.  Recently, the more precise
HERA data has been incorporated into two updated standard sets of
distributions, MRST2001~\cite{MRST2001} and CTEQ6M~\cite{CTEQ6M}.
However, neither set has a sizable change from its predecessor in the
quark and gluon distributions for $x$ in the relevant range 0.01---0.1
at $Q^2 = 10^4$~GeV$^2$.  The new MRST2001 distribution uses a
slightly larger $\alphas(m_Z)=0.119$, which may increase the
importance of the gluon fusion subprocess relative to the $q\bar{q}$
subprocess by a few percent, but overall the effect on the
$\gamma\gamma$ background should be fairly small.

\begin{table}
\caption{\label{Ktable} NLO QCD $K$ factors for $\gamma\gamma$ Higgs signal 
and gluon fusion background. 
Both LO and NLO cross sections are computed using NLO parton distributions.}
\begin{ruledtabular}
\begin{tabular}{lcr}
$\mgg$~(GeV) & $K_{\rm Higgs}$ & $K_{gg\to\gamma\gamma}$
\\ \hline
\hskip5pt 98 & 2.92 & 1.82 \\
         118 & 2.54 & 1.61 \\
         138 & 2.39 & 1.55 \\
\end{tabular}
\end{ruledtabular}
\end{table}

\begin{figure*}[ht]
\includegraphics[width=8.65cm]{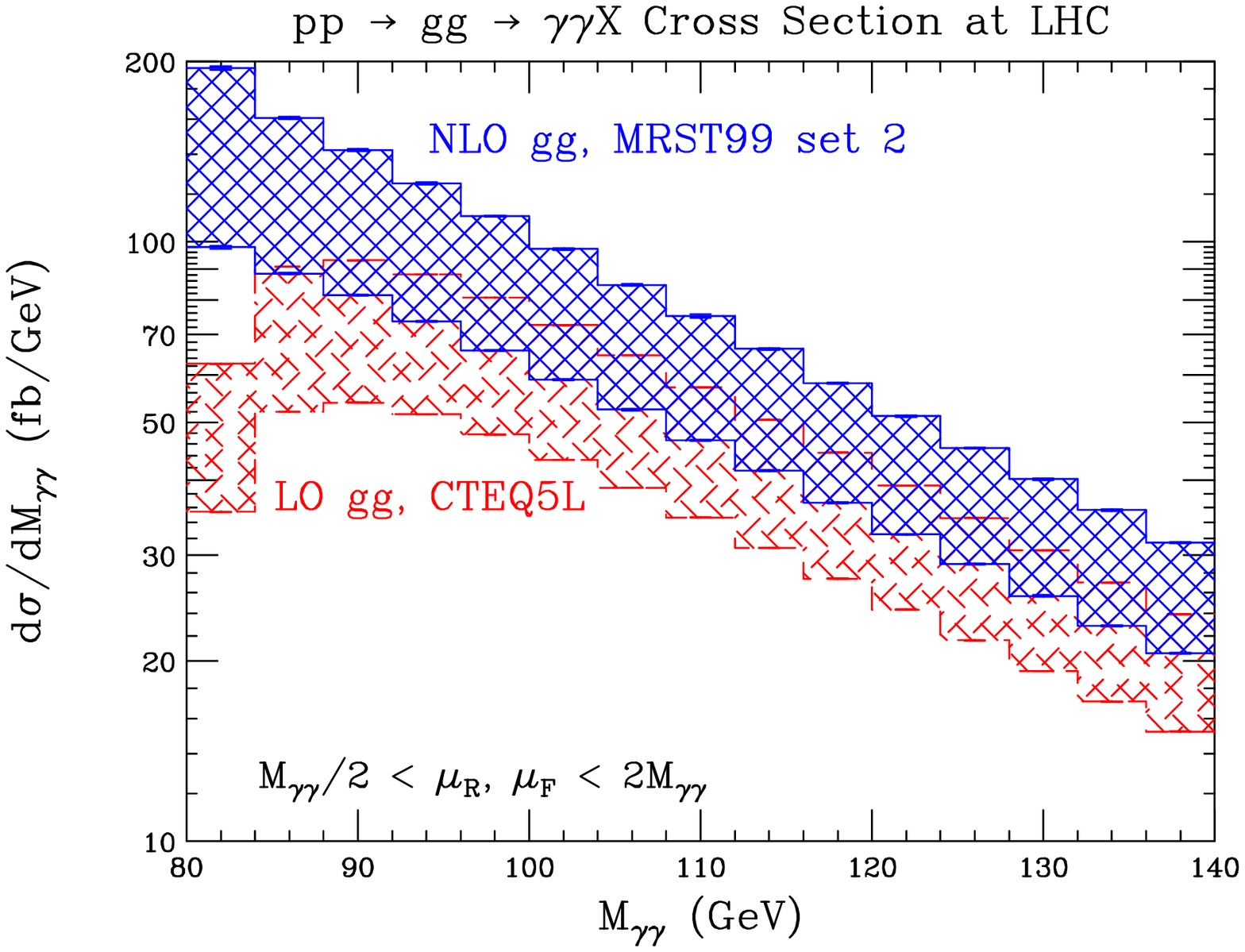}
\includegraphics[width=8.65cm]{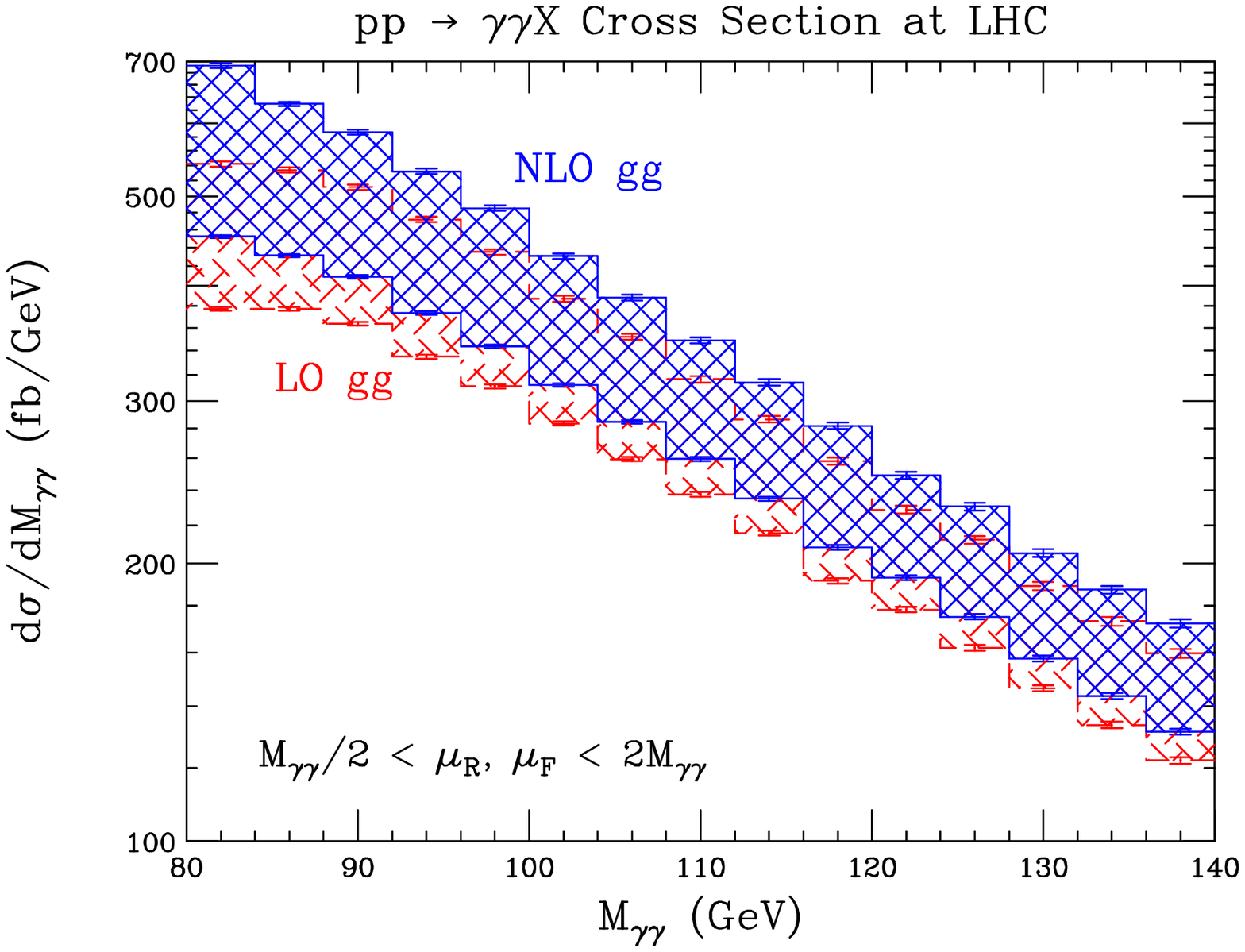}
\caption{\label{fig34_rp4} Scale dependence of (a) the gluon fusion
subprocess contribution to $pp \to \gamma\gamma X$, and (b) the total 
$pp\to \gamma\gamma X$ production cross section, for standard photon
isolation with $R=0.4$, $\etmax = 15$~GeV.  In both plots, the
bands represent the result of varying $\mu_R$ and $\mu_F$ over the square
region $0.5 \mgg < \mu_R , \mu_F < 2 \mgg$.  The dashed (solid) hatched
band corresponds to including the gluon fusion subprocess at LO (NLO).
For the leading order band in (a) only, the LO CTEQ5L parton distributions
were used; otherwise the NLO MRST99 set 2 distributions were employed.}
\end{figure*}

\begin{figure*}[ht]
\includegraphics[width=8.65cm]{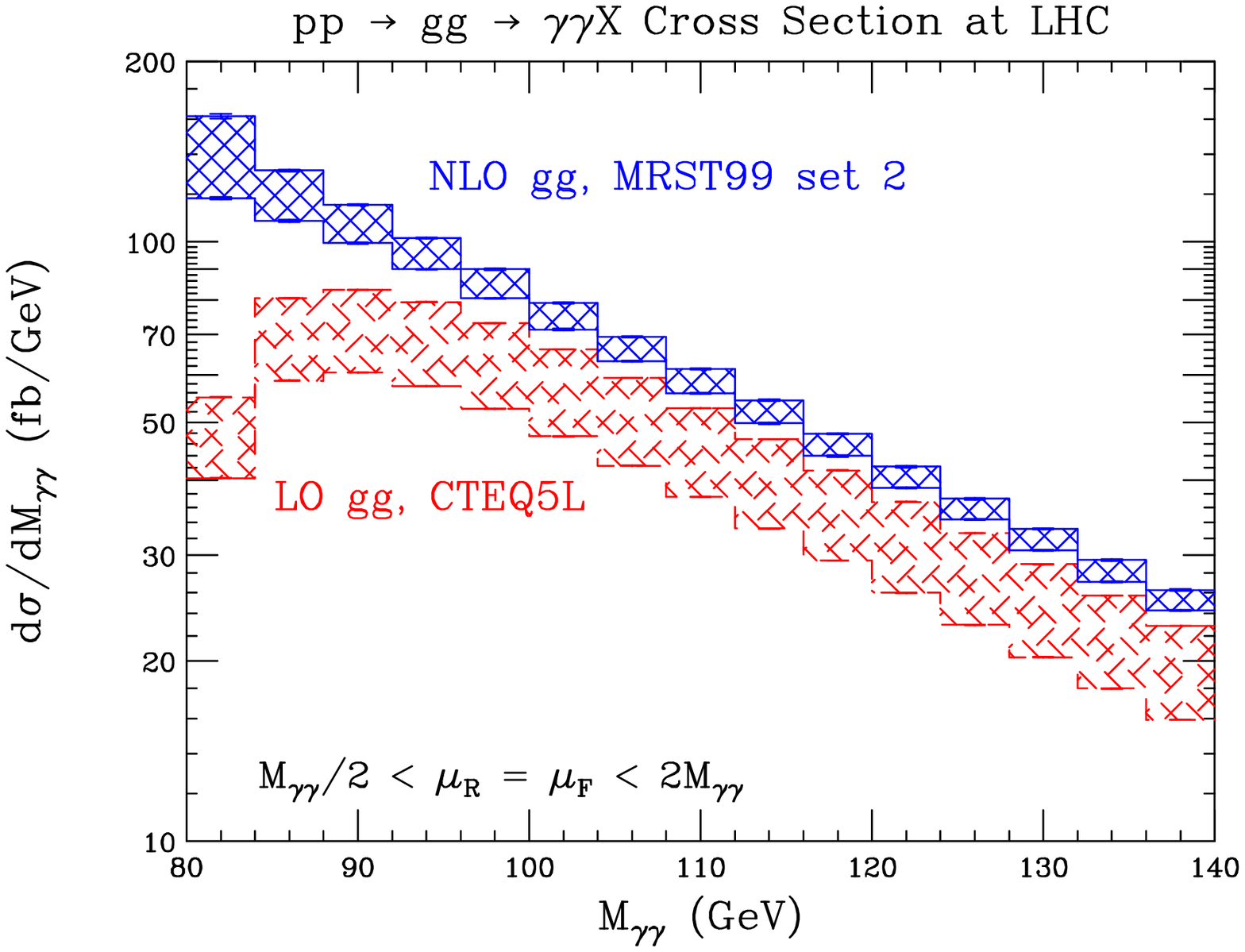}
\includegraphics[width=8.65cm]{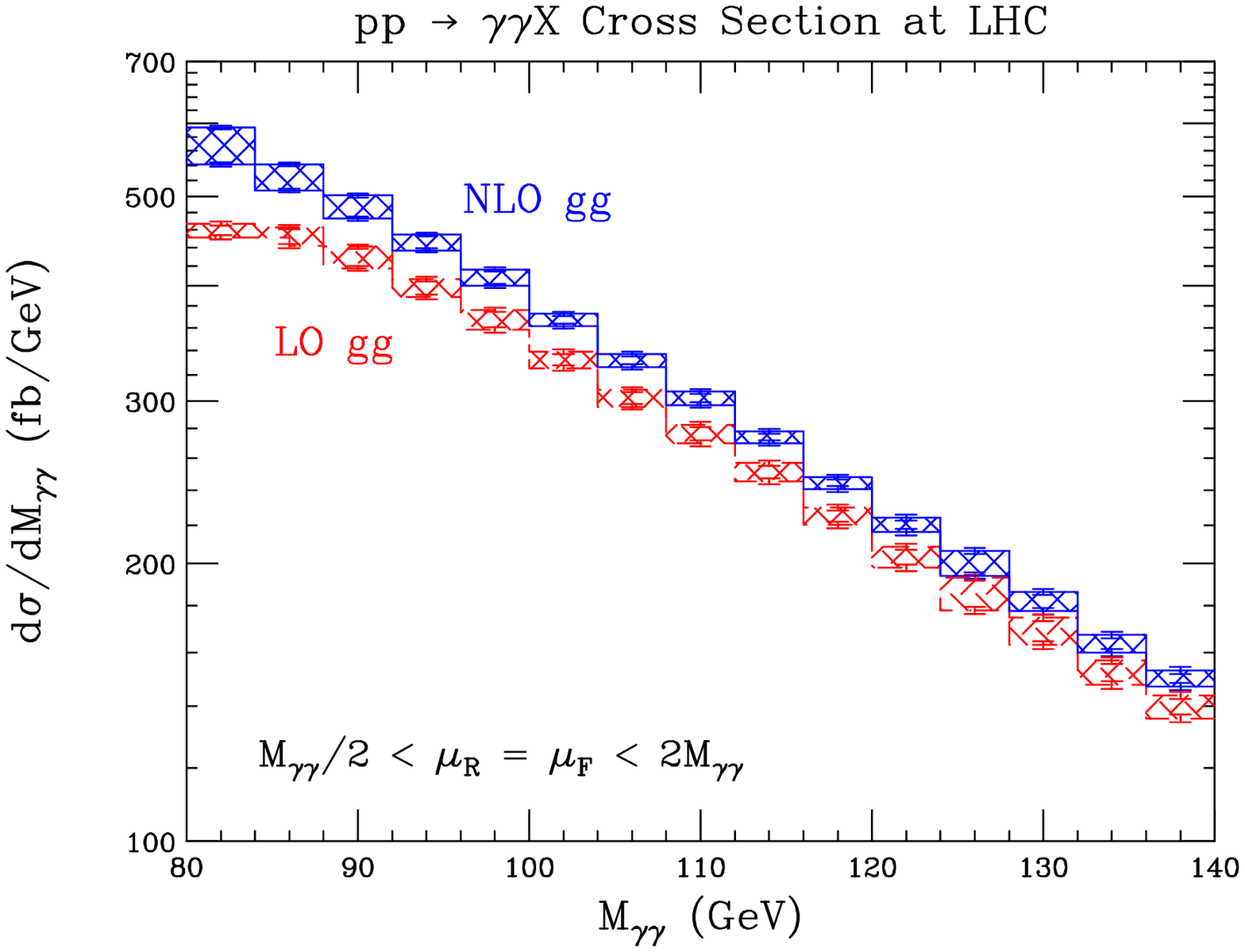}
\caption{\label{fig34_rp4_diag} Scale dependence of (a) the gluon
fusion subprocess contribution to $pp \to \gamma\gamma X$, and (b) the
total $pp\to \gamma\gamma X$ production cross section.  The cuts are
identical to the ones in \fig{fig34_rp4}. The parameters are identical
to the ones used in the corresponding plots in \fig{fig34_rp4}, except
that the renormalization and factorization scales are set equal and
varied between $0.5 \mgg < \mu_R = \mu_F < 2 \mgg$. }
\end{figure*}

In the absence of an NLO calculation, some experimental studies had
used the $K$ factor (ratio of NLO over LO cross section) for Higgs
production by gluon fusion
as an estimate of the $K$ factor for $gg\to\gamma\gamma$.
For example, $K_{\rm Higgs}^{\rm NLO} = 1.85$ was used for a 100~GeV
mass Higgs boson~\cite{CMS}.  The reasoning is that both $gg\to H$ and
$gg\to\gamma\gamma$ involve production of a colorless system from a
gluon-gluon initial state.  One difference between the two processes,
however, is that the $Hgg$ coupling receives a fairly large
short-distance renormalization~\cite{NLOHiggs}, from the top mass
scale,
\begin{equation}
K_{\rm Higgs}^{\rm s.d.} \equiv 
1 + {11\over2} {\alphas\over\pi} \approx 1.2,
\label{Ksd}
\end{equation}
which has no counterpart in the $gg\to\gamma\gamma$ correction.
Another difference stems from the different loop momentum scales
appearing in the real emission diagrams.  For the $gg \rightarrow H g$
case, the momentum in the loop is dominated by the heavy top quark
mass, which is taken to be infinite in our calculation, while for the
$gg \rightarrow \gamma \gamma g$ process the quarks in the loop are
taken to be massless, so that the dominant momentum in the loop is
that of the photons and the emitted gluon.  As a result, the cross
section for $gg \rightarrow \gamma \gamma g$ falls off more quickly
with the emitted gluon transverse momentum than that for $gg
\rightarrow H g$, resulting in a smaller real-emission contribution to
the total NLO cross section.

In comparing signal and background $K$ factors, it is of course useful
to impose the same set of cuts on the photons in each case.  In
table~\ref{Ktable} we list $K$ factors for both the $pp \to gg \to HX
\to \gamma\gamma X$ Higgs production cross section and the $pp \to gg
\to \gamma\gamma X$ gluon fusion background, for three representative
choices of Higgs mass.  (To be precise, the Higgs $K$ factor includes
the subprocesses $qg \to Hq$ and $q\bar q \to Hg$ at NLO; removing
them decreases $K_{\rm Higgs}$ by roughly 5\% at $m_H=$ 118 GeV.)  We
take $\mu_R = \mu_F = 0.5 \mgg$ and impose the same photon acceptance
and isolation cuts as in \fig{fig12_rp4}(a), for both signal and
background.  (With both sets of cuts removed, each $K$ factor is about
10\% smaller at $\mgg=$ 118 GeV, but their ratio is stable to a few
percent.)  We define the ``LO cross section'' entering the $K$ factor
using NLO, rather than LO, parton distributions.  This convention
results in larger $K$ factors than the more standard convention, as
can easily be seen by comparing the LO cross sections using the
CTEQ5M1 and CTEQ5L distributions in \fig{fig12_rp4}(a).  In any case,
the important point is that the $K$ factors for the gluon fusion
component of the di-photon background are significantly smaller than
the $K$ factors for the Higgs signal, even after accounting for the
short-distance contribution~(\ref{Ksd}) to the latter.  This
difference appears to be due to the relatively smaller real-emission
contribution to the background.

In \fig{fig12_rp4}(b) the effects of computing the gluon fusion
subprocess at NLO are shown, for the total NLO $pp \to \gamma\gamma X$
production rate, {\it i.e.} including also the $q\qb\to\gamma\gamma$
and fragmentation contributions at NLO obtained from {\tt
DIPHOX}~\cite{DIPHOX}. As in \fig{fig12_rp4}(a), $\mu_R$ = $\mu_F =
\mgg/2$ and the isolation criterion $R=0.4$, $\etmax = 15$~GeV is
used.  (The lower histogram in \fig{fig12_rp4}(b), 
where the gluon fusion subprocess is treated at LO,
corresponds to the result in fig.~13 of ref.~\cite{DIPHOX}, except
that $R=0.4$, $\etmax = 5$~GeV is used in that plot.)  The increase in
the total irreducible $\gamma\gamma$ background which results from
replacing the LO gluon fusion quark box by the NLO computation is a
modest one, except at the lowest invariant masses relevant only for
non-Standard-Model Higgs searches.  In \fig{fig12_rp4}(b), the
increase ranges from 27\% at $\mgg = 82$~GeV, to 10\% at 100~GeV, and
only 6\% at 138~GeV.  For the most interesting mass range for 
the Higgs boson in this channel, 115 GeV $< m_H <$ 140 GeV, the overall
effect on the square root of the background is under 5\%.
The larger increase at smaller $\mgg$ simply
reflects the fact that the LO contribution vanishes, due to the
kinematic cuts, as $\mgg \to 80$ GeV.  This feature is seen most
visibly in \fig{fig12_rp4}(a).

In \fig{fig34_rp4}, the dependence of the $\gamma\gamma$ background on
the renormalization scale $\mu_R$ and factorization scale $\mu_F$ is
illustrated by varying them independently over the square region $0.5
\mgg < \mu_R , \mu_F < 2 \mgg$.  \Fig{fig34_rp4}(a) shows the
variation for the gluon fusion subprocess contribution alone, while
\fig{fig34_rp4}(b) shows the variation for the total production rate,
treating the $q\qb\to\gamma\gamma$ and fragmentation contributions at
NLO.  The same photon isolation criterion is used as in
\fig{fig12_rp4}.  In \fig{fig34_rp4}(a), the leading-order (dashed
hatched) band is computed using the LO parton distribution
CTEQ5L, which is a bit more appropriate when considering
this subprocess in isolation.  In all cases, the maximum cross section
in the band at a given $\mgg$ comes from setting $\mu_R = 0.5 \mgg$
and $\mu_F = 2 \mgg$, while the minimum cross section comes from
setting $\mu_R = 2 \mgg$ and $\mu_F = 0.5 \mgg$.  Allowing independent
variations for $\mu_R$ and $\mu_F$ results in NLO bands which are not
appreciably narrower than the LO bands.  In contrast, varying $\mu_R$
and $\mu_F$ together, {\it i.e.}  $\mu_R = \mu_F = \chi \mgg$ with
$0.5 < \chi < 2$, as is more conventional, leads to much less scale
variation for the gluon fusion subprocess at NLO than at LO, as shown
in \fig{fig34_rp4_diag}.  (These general features are also 
qualitatively present in the Higgs production cross section as well, 
although the larger NLO $K$ factor leads to stronger renormalization scale
dependence in that case; see {\it e.g.} ref.~\cite{CdFG}.)
The considerable improvement in the scale variation for the $gg$ 
contribution depicted in \fig{fig34_rp4_diag}(a)
is diluted in \fig{fig34_rp4_diag}(b), where the contributions with 
quark initial states and fragmentation are added in.

In conclusion, the NLO corrections to the $gg\to\gamma\gamma$ subprocess
have a modest effect on the total irreducible di-photon background
to the Higgs search.  Thus this subprocess can be considered to be 
under adequate theoretical control.


\section{Statistical Significance of Higgs Signal}
\label{StatisticsSection}

In this section we investigate the kinematic features of the Higgs
signal and background, starting with photon isolation criteria.  To
facilitate this study we consider a crude approximation to an
experimental analysis at the LHC.  We assume a Higgs mass of 118 GeV,
and we count the number of events in mass bins of 4 GeV for $30$
fb$^{-1}$ of integrated luminosity, corresponding to 3 years of
running at low luminosity, ${\cal L}=10^{33} $cm$^{-2}$s$^{-1}$.  We
note that this choice of mass bin is slightly larger than the
optimized mass bins of 2.74 and 3.44 GeV used in the ATLAS
study of ref.~\cite{Tisserand}.  We also include an efficiency factor
of 0.57 for both signal and background, corresponding to the
combination of 0.81 per $\gamma$ for $\gamma$/jet identification and
0.87 for fiducial cuts (mainly the transition between barrel and
endcap) found in that analysis.  Finally, we include a reducible
background of 20\% of the $\gamma\gamma$ continuum background, which
we assume is possible after the $\gamma$/jet
identification~\cite{Tisserand}.  For this analysis we take the
efficiency factors and the percentage of reducible background as
independent of the isolation cuts; to investigate this further would
require a more serious experimental analysis, beyond the scope of this
work.

In computing the statistical significance we ignore interference
between the Higgs signal and the background.  In the Standard Model,
the interference terms are on the order of a few percent of the Higgs
signal~\cite{gHgamInterf}, and do not significantly alter any of our
conclusions.  The small size of the interference is due mostly to the
extreme sharpness of the underlying Higgs resonance which, before
smearing with the detector resolution, gives a peak in the cross
section rising about a factor of a hundred over the background. Hence
the interference contribution should not be more than about 20 percent
of the signal.  However, it is less than this because the primary
Higgs production mode is via gluon fusion, so only the $gg \to
\gamma\gamma$ component of the background can interfere.  Also,
because the experimental width is much greater than the intrinsic
width, only the integral in $M_{\gamma\gamma}$ across the lineshape is
observable.  This integral vanishes unless there is a relative phase
between the production ($gg \to H$), decay ($H \to \gamma\gamma$), and
background ($gg \to \gamma\gamma$) amplitudes.  The phase happens to
vanish, up to small quark mass effects, when the background amplitude
(for identical-helicity photons) is evaluated at one loop.  The extra
power of $\alpha_s$ in the two-loop amplitude then provides an
additional suppression factor.

\subsection{Effects of varying photon isolation criteria
\label{IsolationSubsection}}

We first consider the effects on both signal and background
of varying the photon isolation criteria, before turning in
section~\ref{JetVetoSubsection} to the
phenomenologically more viable method of using a jet veto to impose
more stringent cuts.
As mentioned in the introduction,
photon isolation can be achieved by either a standard or a smooth cone
criterion. In section~\ref{TotalCrossSubsection} we presented cross
section results for the standard cone criterion with $R = 0.4$ and
$\etmax = 15$~GeV, values typical to previous analyses.  Now we shall
investigate how the $\gamma\gamma$ background varies, relative to the
$H\to\gamma\gamma$ signal, as we change the isolation criteria.  In
particular, we would like to determine the parton-level statistical
significance of the signal as a function of photon isolation.

\begin{figure*}
\includegraphics[width=8.65cm]{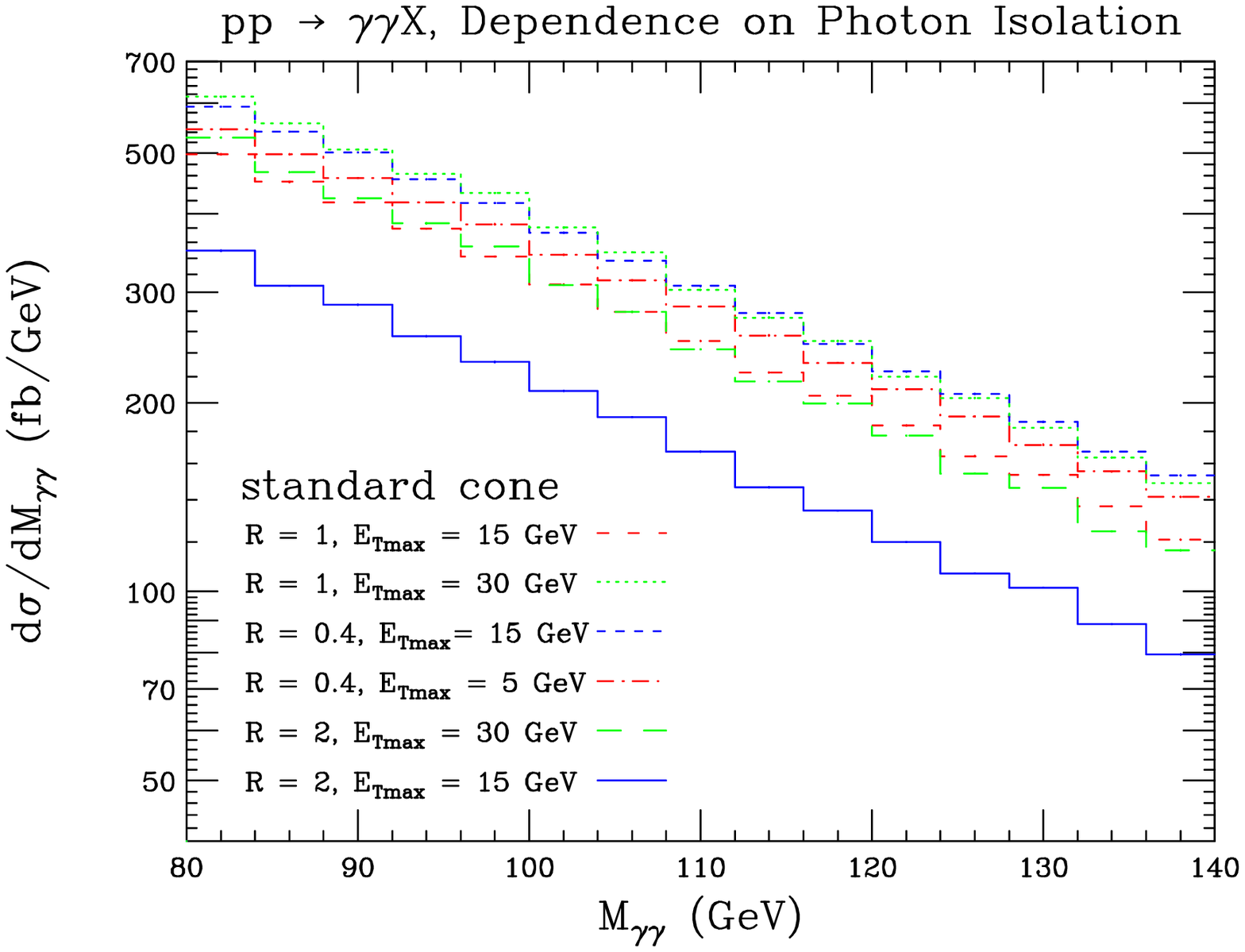}
\includegraphics[width=8.65cm]{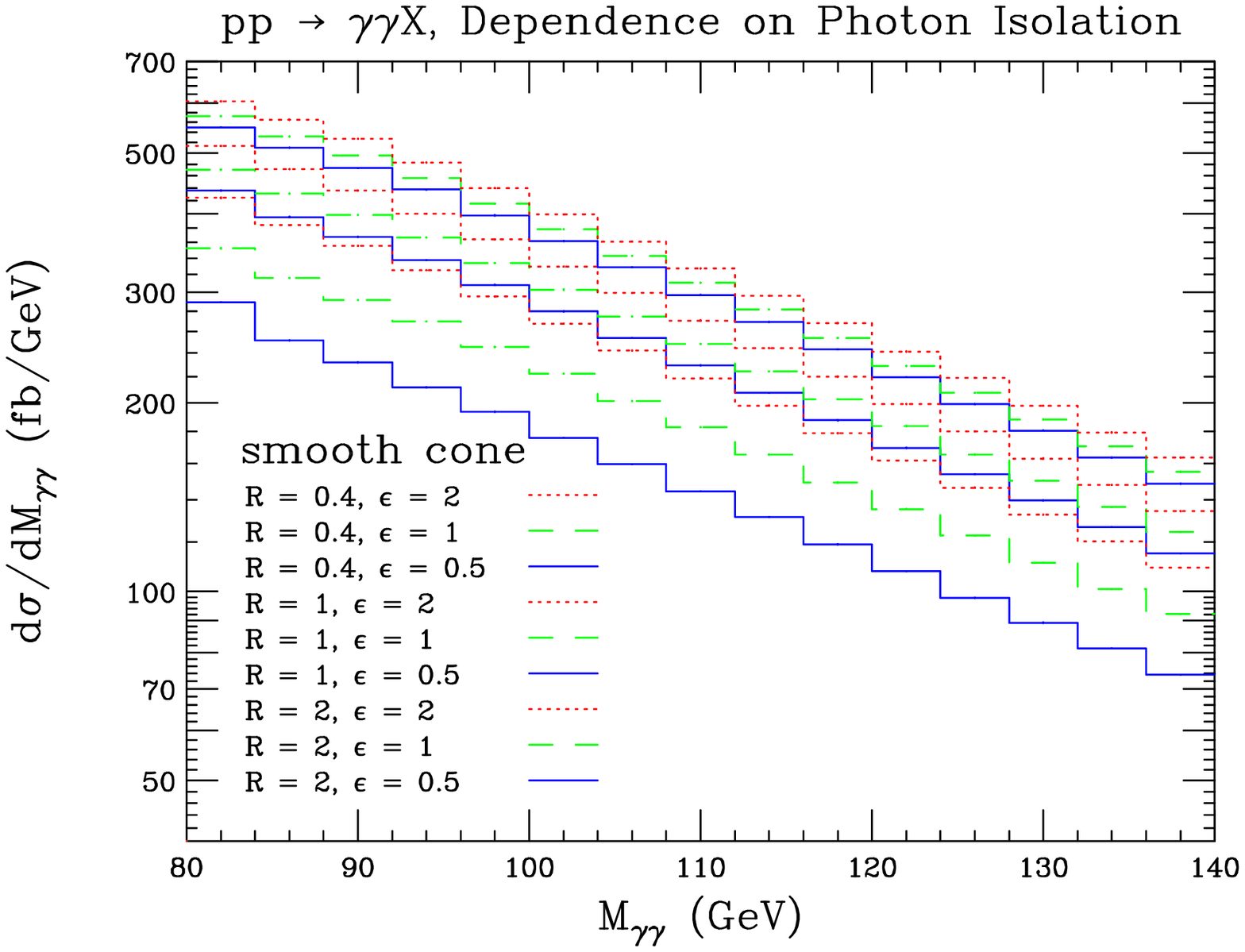}
\caption{\label{fig56} Dependence of $pp\to\gamma\gamma X$ at the LHC 
on photon isolation cuts, for (a) a set of standard cone isolation
parameters, $R$ and $\etmax$, and (b) a set of smooth cone isolation
parameters, $R$ and $\eps$.  All plots are for MRST99 set 2 partons, and 
$\mu_R = \mu_F = 0.5 \mgg$. }
\end{figure*}

\Fig{fig56}(a) shows how the $pp\to\gamma\gamma X$ production rate at 
the LHC depends on the parameters $R$ and $\etmax$ of the standard cone
isolation definition, while \Fig{fig56}(b) presents analogous information
for the smooth cone criterion.  As the isolation becomes more severe, 
{\it i.e.}
$R$ is increased or $\etmax$ or $\epsilon$ are decreased, the direct 
$pp\to\gamma\gamma X$ background becomes more suppressed.   The 
large sensitivity to these parameters is indicative of the $q\gamma$ collinear 
singularity in the NLO $q\bar q \to \gamma\gamma X$ cross section.
Since the QCD radiation in Higgs production has no such singularity, it should
have no correlation with the photon directions, and therefore it should be
less sensitive to the isolation criterion.

\begin{figure*}
\includegraphics[width=8.65cm]{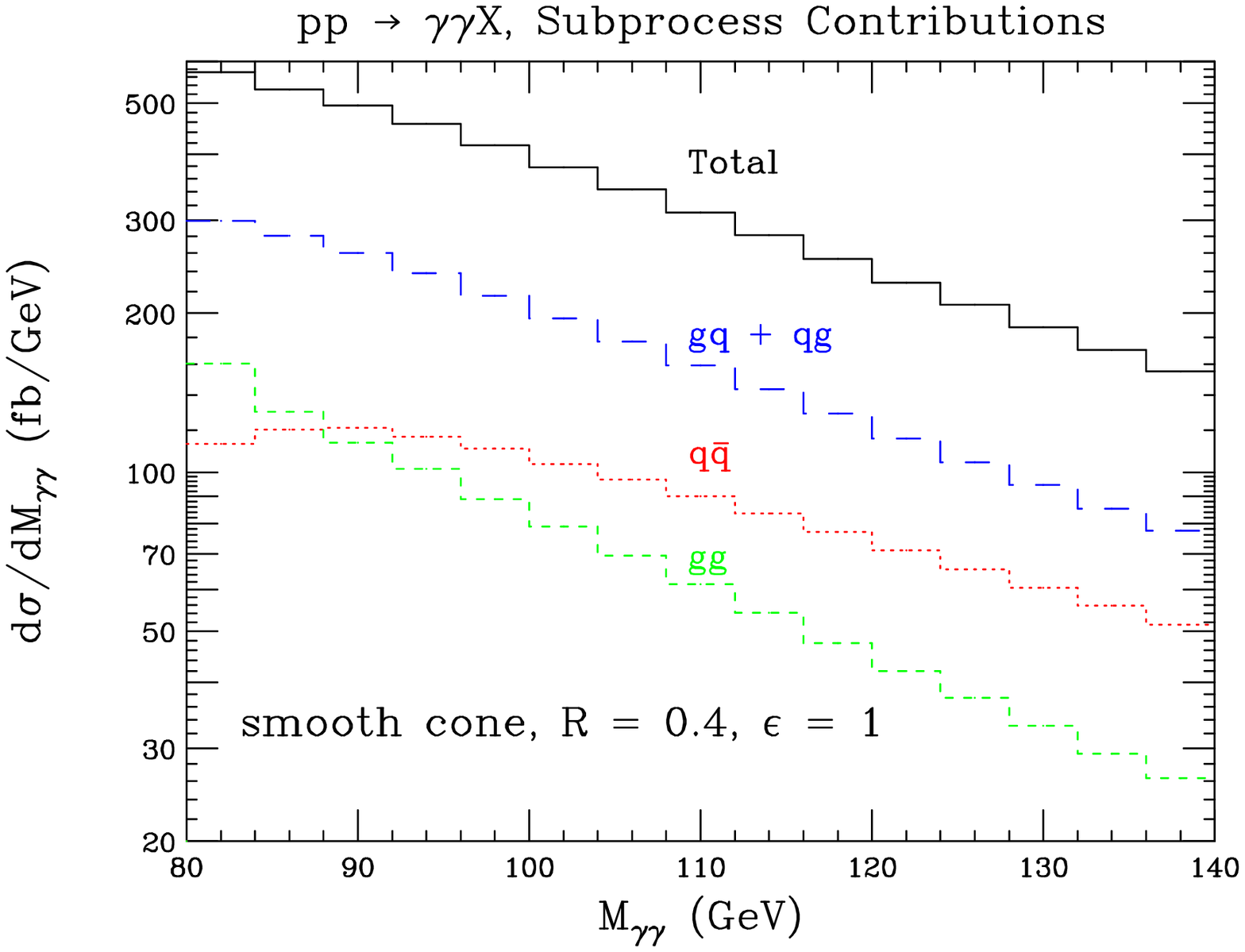}
\includegraphics[width=8.65cm]{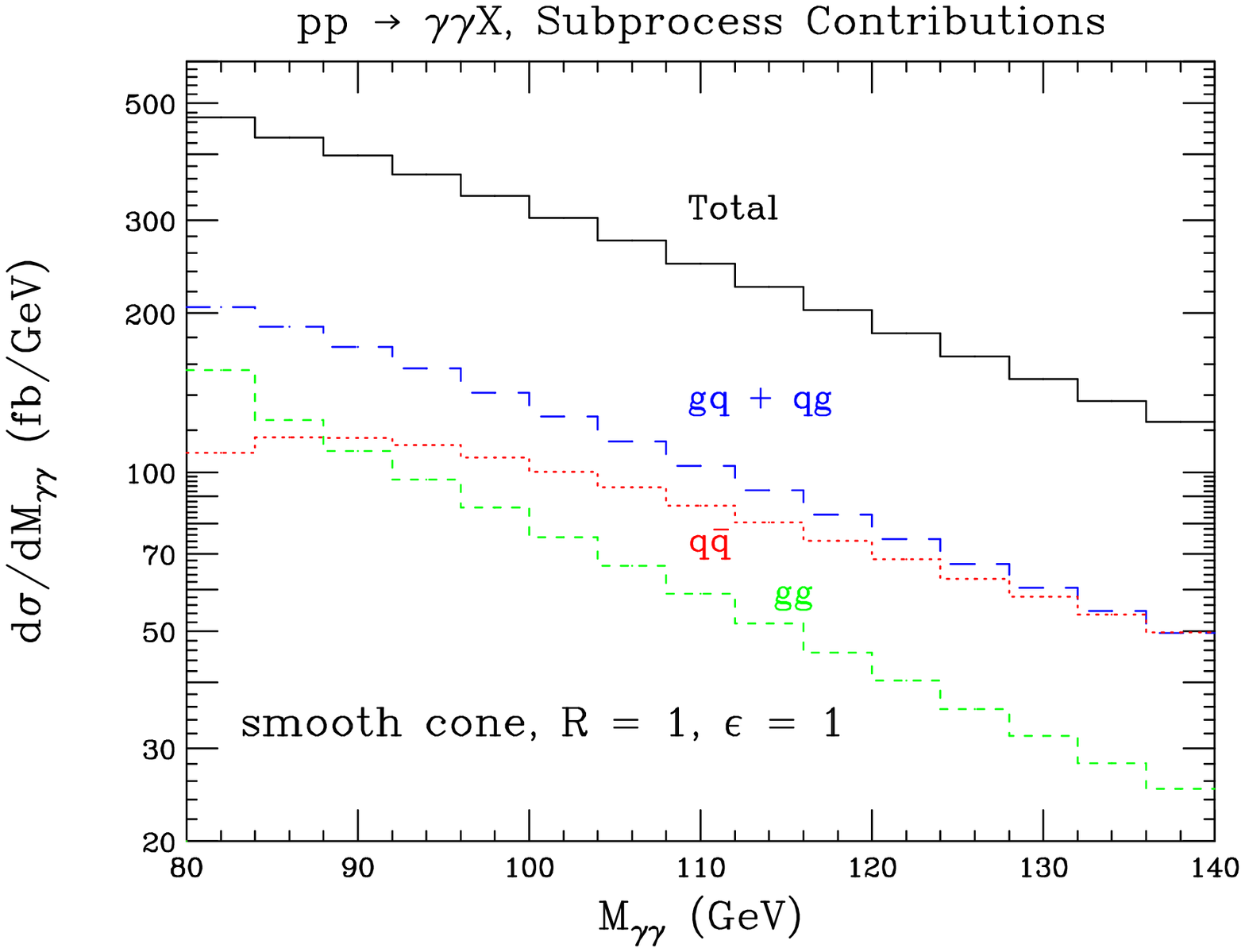}
\caption{\label{Fig6components} Contributions of the different subprocesses to
$pp\to\gamma\gamma X$ at the LHC, as a function of di-photon invariant
mass, for two choices of smooth cone isolation parameters:
(a) $R = 0.4$, $\eps=1$, and (b) $R = 1$, $\eps=1$.  
The plots are computed at NLO, for MRST99 set 2 partons, with acceptance
cuts~(\ref{AccCuts}), and $\mu_R = \mu_F = 0.5 \mgg$.  }
\end{figure*}

To see this more clearly, it is instructive to plot individually the
various subprocess contributions to the $\gamma\gamma$ background.
\Fig{Fig6components}(a) and (b), plotted for two different smooth cone
isolation criteria, show how the $qg$ component is reduced relative to
$q\qb$ and $gg$ as the isolation requirement is made more severe.  For
instance, in the bin centered at $M_{\gamma\gamma}=118$ GeV, the $qg$
component decreases by 36\%, in going from $R=0.4,\ \epsilon=1$ to
$R=1,\ \epsilon=1$, while the $q\qb$ and $gg$ components each only
decrease by about 4\%.  The smooth cone criterion was used to simplify
the discussion, since there are no fragmentation contributions;
however, the results are qualitatively similar for the standard cone
isolation.

We give the number of Higgs signal ($S$) and background ($B$) events
and the statistical significance ($S/\sqrt{B}$) in the bin 116 GeV
$<M_{\gamma\gamma}<$ 120 GeV for the different choices of standard
cone parameters in table~\ref{SBstandardtable}, and for the choices of
smooth cone parameters in table~\ref{SBsmoothtable}.  For the standard
cone, we find a statistical significance of 7.3 at $R=1,\ \etmax=15$
GeV, for a modest gain of about 4\% over the value of 7.0 at $R=0.4,\
\etmax=15$ GeV.  The significance can be increased further to 7.5 by
the very severe cut of $R=2,\ \etmax=15$ GeV, for a gain of 7\% over
the value at $R=0.4,\ \etmax=15$ GeV.  For the smooth cone, the
statistical significance appears to have a maximum of about 7.4 for
$R=1,\ \epsilon=0.5$ and $R=2,\ \epsilon=2$.  It is clear from these
results that for either the smooth or standard cones the statistical
significance depends on the isolation cuts only rather weakly.

\begin{table}
\caption{\label{SBstandardtable} Number of signal and background events
and the statistical significance for $pp \to HX \to \gamma\gamma X$ in a
bin 116 GeV $<M_{\gamma\gamma}<$ 120 GeV for $30$ fb$^{-1}$ integrated
luminosity using standard cone isolation.  The Higgs mass is taken to be
$m_H=$ 118 GeV.  Other experimental assumptions are given in the text. }
\begin{ruledtabular}
\begin{tabular}{cccc}
$(R,\etmax$ in GeV) & $S$ & $B$ & $S/\sqrt{B}$
\\ \hline
         (0.4,15) & 993 & 20,400 & 7.0 \\ 
         (0.4,5) & 980 & 19,000 & 7.1 \\ 
         (1,30) & 979 & 20,600 & 6.8 \\ 
         (1,15) & 952 & 16,900 & 7.3 \\ 
         (2,30) & 896 & 16,400 & 7.0 \\ 
         (2,15) & 789 & 11,000 & 7.5 \\ 
\end{tabular}
\end{ruledtabular}
\end{table}

\begin{table}
\caption{\label{SBsmoothtable} Number of signal and background events and
the statistical significance for $pp \to HX \to \gamma\gamma X$ in a bin
116 GeV $<M_{\gamma\gamma}<$ 120 GeV for $30$ fb$^{-1}$ integrated
luminosity using smooth cone isolation.  The Higgs mass is taken to be
$m_H=$ 118 GeV.  Other experimental assumptions are given in the text.}
\begin{ruledtabular}
\begin{tabular}{cccc}
$(R,\epsilon)$ & $S$ & $B$ & $S/\sqrt{B}$
\\ \hline
(0.4,2) &    993   &  22,000   &     6.7 \\
(0.4,1) &    992   & 20,800    &    6.9  \\
(0.4,0.5) &  985   & 20,000    &    7.0  \\
(1,2)  &     969   & 18,100    &    7.2  \\
(1,1)  &     948   & 16,700    &    7.3  \\
(1,.5) &     915   & 15,400    &    7.4  \\
(2,2)  &     893   & 14,700    &    7.4  \\
(2,1)  &     806   & 12,300    &    7.3  \\
(2,.5) &     685   & \hskip .17 cm  9,800    &    6.9  \\
\end{tabular}
\end{ruledtabular}
\end{table}

The smallest cone size, $R=0.4$, used in tables~\ref{SBstandardtable}
and~\ref{SBsmoothtable}, is the standard cone size used in previous
studies. Recently it was observed~\cite{CataniPhotons} that logarithms
of the form $\alpha_s(\mu) \ln(1/R^2)$~\cite{GV} can invalidate an NLO
calculation of prompt photon production: For $R=0.1$ the NLO
single-photon cross section with isolation was larger than the cross
section with no isolation, which is clearly an unphysical result.  It
is not yet known precisely how small $R$ can be taken before the NLO
calculation begins to break down.  However, for $R=0.4$, the
$\ln(1/R^2)$ factor is 2.5 times smaller than it is for the
pathological case of $R=0.1$.  Also, the physical scale in the
di-photon invariant masses relevant for the LHC Higgs search is
significantly higher than the single-photon $\pt(\gamma) = 15$ GeV
case studied in ref.~\cite{CataniPhotons}, rendering $\alpha_s(\mu)$
smaller as well.  Finally, we note that the large logarithms do not
arise from the gluon fusion contributions, because there is no $g
\gamma$ collinear singularity.  Hence the effect of the logarithms in
the single-photon case (where gluon fusion is not important and was
not included) is diluted somewhat in the di-photon case by the gluon
fusion contribution.  Nevertheless, further study of this situation,
including possibly resummation of the logarithms, could be helpful.

A more critical issue for this analysis is that the most severe
isolation parameters may not be phenomenologically viable, for both
the theoretical and experimental reasons mentioned in the
introduction. Theoretically, it is not infrared safe to forbid all
gluon radiation into any finite region of phase space. If the
isolation criteria approach this limit, the perturbative predictions
become subject to large corrections and therefore become
unreliable~\cite{GV,CataniPhotons}.  After all, two $R=2$ cones can
cover most of the $(\eta,\phi)$ plane within the detector acceptance,
and $\etmax=15$ GeV is not a lot of energy at the LHC.  On the
experimental side, the efficiency for collecting signal events may
decline for reasons that are absent from the NLO Monte Carlo.
Instrumental (calorimeter) noise, pile up, and energy deposition from
the underlying event plus overlapping minimum bias collisions, all
contribute an average energy in a cone which scales roughly as the
area of the cone.  Thus one might expect that when $R$ is increased,
one should also increase $\etmax$, roughly like $R^2$.  For the
$R=0.4$ cone typically used, pile up and underlying events start to
saturate the cone at $\etmax \approx 2.5$ GeV~\cite{DIPHOX,Wielers}.
For $R=2$, saturation would most likely be occurring at $\etmax = 15$
GeV, perhaps even at 30 GeV.  Another potential problem is that as one
varies the cuts to reduce the irreducible contributions, one must be
sure that the reducible contributions do not get larger, undoing the
improvement.  For example, we have raised the total transverse energy
allowed near the photon, in going from $R=0.4$, $\etmax=15$ GeV to,
say, $R=2$, $\etmax=30$ GeV, and this may allow more of the reducible
background to enter.  This question could be addressed by studies
along the lines of ref.~\cite{PiBkgd}.  In any case, it is clear that
a phenomenologically more sensible method for rejecting events with
hadronic energy near the photons is required.


\subsection{Jet veto}
\label{JetVetoSubsection}

As mentioned in the introduction, a veto on nearby jets offers another
way to suppress the QCD background, in particular the 
$qg \to \gamma\gamma q$ process~\cite{Tisserand}.
At the NLO parton level, at least for direct processes, it corresponds 
closely to increasing the size of the cone.  However, because transverse 
energy is being forbidden into a smaller area (the jet cone size), for
the same amount of suppression at NLO, the jet veto is a more infrared safe
criterion, and it should also have better experimental properties
(less loss of signal due to noise, overlapping events, {\it etc.}).

Jet vetoes have been considered previously in search strategies for
other Higgs decay modes, particularly $H \to W^+ W^- \to e^\pm \mu^\mp
\ptmiss$~\cite{RunIIExpectations,CMS2,ATLAS,DD,HanZhangCdFG}.  In
those cases, typically a general veto is applied on all jets in the
detector acceptance with $E_T$ above a certain value.  For the
$\gamma\gamma$ mode, we would only like to veto on jets ``close'' to
the photon candidates.  Such nearby jets are more likely to come from
the $qg\to \gamma\gamma q$ subprocess, because of the final state
$q\gamma$ collinear singularity, than from the Higgs production
process $gg \to H X$.  On the other hand, because the gluon is in a
larger color representation than the quark, $gg$-initiated production
of a color singlet object tends to be jettier overall than production
initiated by $q\bar{q}$ or $qg$ initial states.  In fact, cuts
requiring a minimum transverse momentum of the $\gamma\gamma$ pair,
$Q_T > 30$ GeV, have been proposed to take advantage of this
fact~\cite{Abdullin,GGGamGamGa,GGGamGamGb}, and enhance the signal.
So only jets ``sufficiently'' near a photon candidate should be
vetoed.

\begin{figure}[t]
\includegraphics[width=8.65cm]{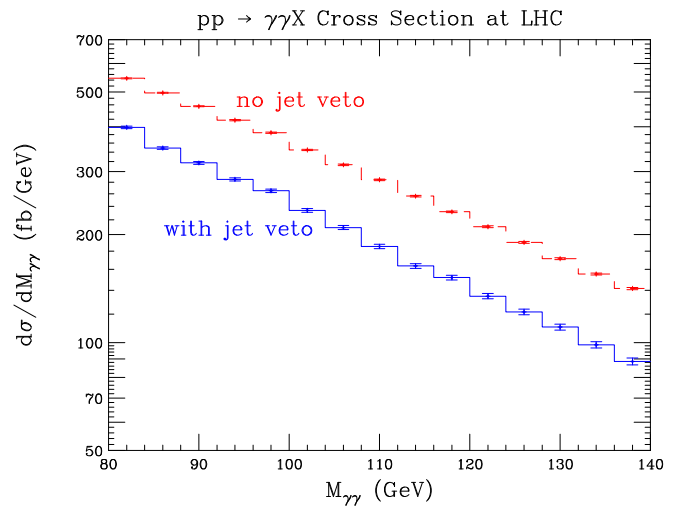}
\caption{\label{fig_jetveto} 
Effect on the background process $pp \to \gamma\gamma X$ of vetoing on 
jets with $\et > 15$ GeV within $\Rjet=2$ of either photon candidate.
The veto is on top of standard cone isolation with $R=0.4$, 
$\etmax = 5$~GeV.  MRST99 set 2 partons are used, with the default
acceptance cuts~(\ref{AccCuts}) and scale choice~(\ref{defaultmu}).}
\end{figure}

We implement the jet veto on top of a standard photon isolation cone,
represented by the inner cone in \fig{figjetveto}.  We require that
there is no jet with a transverse energy $\et > \etjet$ within a
radius $\Rjet$ of the photon, represented by the outer cone in
\fig{figjetveto}. We do not include hadronic energy inside the inner
cone in defining this jet.  Then the results at the NLO parton level
do not depend on the cone size $R^{\rm cone}_{\rm jet}$ used in the
jet algorithm, but for definiteness we suppose $R^{\rm cone}_{\rm jet}
= 0.7$.

As an example of the jet veto suppression, \Fig{fig_jetveto} shows the
background suppression obtained for a jet veto using $\Rjet=2$ and
$\etjet = 15$ GeV on top of a standard isolation cone with $R=0.4$ and
$\etmax=5$ GeV.  For this standard isolation cone the fragmentation
contribution is rather small, amounting to about 10 percent of the
total.  This simplifies the calculation of the jet veto since we can
ignore the action of the jet veto on the small fragmentation part.
For the direct piece at NLO, a jet to be vetoed amounts to a lone
parton with transverse energy $E_T > \etjet = 15$ GeV between the
inner and outer cones $0.4 < R < 2$.  By ignoring the jet veto
rejection of the fragmentation term, the background is overestimated
by a few percent.  In this approximation, with $m_H = 118$ GeV, the
bin 116 GeV $<M_{\gamma\gamma}<$ 120 GeV has 776 signal events and
12,600 background events, leading to a statistical significance of
$S/\sqrt{B} = 6.9$. Even though the background drops from 19,000
events for the standard isolation case with $R=0.4$ and $\etmax=5$ GeV
to 12,600 events when the jet veto is included, the statistical
significance is essentially unchanged compared to this case.  This
illustrates the rather disappointing insensitivity of the statistical
significance to the presence of the jet veto.


\subsection{Kinematic distributions of signal and background photons
\label{AngularSubsection}}

The situation can be improved somewhat by including information from
the photon angular distribution.  Since the Higgs boson is a scalar, its
decay to two photons is isotropic in its rest frame.  In contrast, the
$\gamma\gamma$ background processes tend to be more peaked toward the
beam axis.  Thus, the angular distribution of the photons can help
separate the signal from the background.

\begin{figure}[t]
\includegraphics[width=8.65cm]{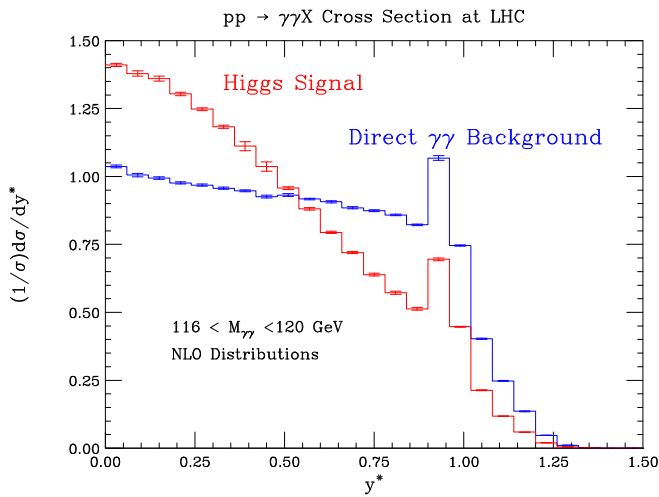}
\caption{\label{fig9} The angular distribution as a function of $y^*$
for 116 GeV $< \mgg <$ 120 GeV.  The renormalization and factorization
scales are $\mu_R = \mu_F = 0.5 \mgg$ and the smooth cone isolation
cuts are $R=1, \epsilon=1$. The Higgs mass is taken to be $m_H = 118$
GeV. The peaks near $y^*=0.94$ occur in a region where
the NLO calculation breaks down and is not trustworthy.}
\end{figure}

\Fig{fig9} shows the normalized distribution in the di-photon rapidity
difference, $y^* = (y(\gamma_1) - y(\gamma_2))/2$.  This variable is
convenient because it is simple to determine experimentally, and at
lowest order it is related to the center-of-mass scattering angle
$\theta^*$ for $q\bar{q} \to \gamma\gamma$ or $gg \to \gamma\gamma$ by
$\cos\theta^* = \tanh y^*$.  The renormalization and factorization
scales are set to our default values~(\ref{defaultmu}), and only
events in the mass bin 116~GeV $< \mgg <$ 120~GeV are kept.  The
smooth cone isolation is used with parameters $R=1$ and $\epsilon=1$;
similar distributions are obtained using a standard cone isolation.
As can be seen in \fig{fig9} the angular distribution of the Higgs
signal events is rather different from the background.  We can
estimate the significance that could be obtained by using a maximum
likelihood function with this information to be
\begin{equation}
\left(\sum_i{S^2_i\over B_i}\right)^{1/2}=7.7\ ,
\label{OptimalSig}
\end{equation}
where the sum is over the bins in $y^*$.  This number is to be 
compared with a significance of
\begin{equation}
{ \sum_i S_i \over ( \sum_i B_i )^{1/2} } = 7.3\ ,
\label{UnOptimalSig}
\end{equation}
without using the angular information.  The 4\% relative improvement in 
significance should also hold roughly for $y^*$ distributions constructed
using a standard cone isolation.

An interesting feature in \fig{fig9} are the peaks in both the signal
and background in the bins near 0.90-1.00.  These peaks are
attributable to a breakdown of the NLO approximation near the LO
kinematic boundary in $y^*$, whose location is dictated by the
$p_T(\gamma_1) > 40$ GeV cut~(\ref{AccCuts}) and $\mgg \approx 118$
GeV.  At LO the two photons are constrained to have vanishing total
transverse momentum $Q_T \equiv |\vec{p}_{1\rm T}+\vec{p}_{2\rm T}|$,
which leads to $y^* < 0.94$.  At NLO, events with a radiated gluon can
have nonzero $Q_T$, which removes the constraint on $y^*$.  For small
$Q_T$, the NLO cross section is very unstable and must be resummed in
$\alpha_s \ln^2(Q_T/p_T)$, as in refs.~\cite{GGGamGamGb,BY}.  Similar
phenomena have been described in earlier work on isolated
photons~\cite{IsolatedPhotonIR,DIPHOX}.  A general description of such
``edges'' has also been given~\cite{CataniWebber}.  In \fig{fig9},
$\ln(Q_T/p_T)$ becomes appreciable only in the two bins centered at
$y^* = 0.93$ and $0.99$.  The bins to the left of $y^* = 0.94$ do not
contain these uncancelled logarithms because the virtual corrections,
with LO kinematics, can contribute and cancel them.  Of course, all
bins to the right of $y^* = 0.94$ are effectively being calculated at
LO, hence their overall normalization is not as trustworthy.

One might be concerned that the parton-level normalized distributions
shown in \fig{fig9} will be distorted by higher-order terms, soft
physics, and detector effects, rendering the information inadequate for
improving the significance of the signal.  However, this is not the
case, because 1) only an {\it approximate} knowledge of the {\it
relative} shapes of signal and background is required to get most of
the benefit, and 2) the background distribution can be measured
experimentally.  Once a putative peak is identified in the $M_{\gamma
\gamma}$ spectrum, the $y^*$ distribution in the sideband regions
above and below the peak can be measured. This information, along with
that in \fig{fig9}, can be used to estimate the true signal
distribution, including detector effects, etc.  One can then apply an
optimal observable or maximum likelihood analysis similar to the one
described above.

\begin{figure}[t]
\includegraphics[width=8.65cm]{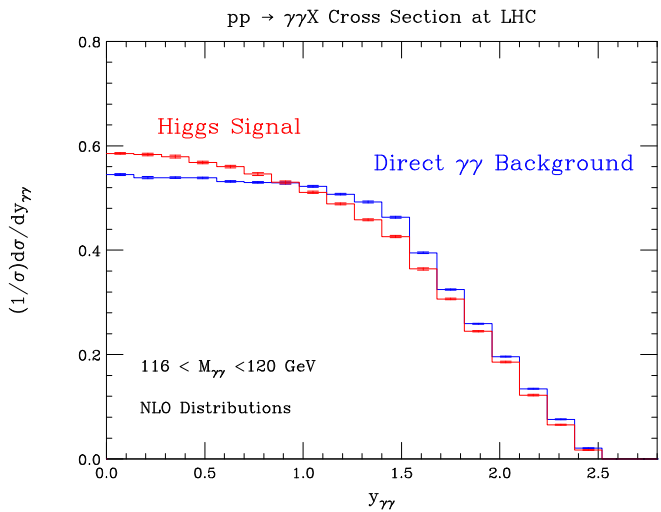}
\caption{\label{fig_ygg} The signal and background distributions as 
a function of $y_{\gamma \gamma}$
for 116 GeV $< \mgg <$ 120 GeV.  The renormalization and factorization
scales are $\mu_R = \mu_F = 0.5 \mgg$ and the smooth cone isolation
cuts are $R=1, \epsilon=1$. The Higgs mass is taken to be $m_H = 118$
GeV. }
\end{figure}

The final distribution that we consider is that of $y_{\gamma\gamma}$, 
defined to be the rapidity of $p_{\gamma\gamma}
=p_{\gamma_1}+p_{\gamma_2}$, the four-vector sum of the two
photon momenta.  For the case of the Higgs signal, this is just the
rapidity of the Higgs boson.  We plot the normalized distribution in
$y_{\gamma \gamma}$ in \Fig{fig_ygg}, for the same choice of mass bin,
isolation cuts, and scale choices as for \Fig{fig9}.  The difference
between the signal and the background distributions can be mostly
attributed to the different parton luminosities involved in the
production; the Higgs signal is produced by a predominantly $gg$
initial state, whereas the di-photon background gets significant
contributions from each of $q\bar q$, $qg$, and $gg$.  In this case
the use of $y_{\gamma \gamma}$ in a maximum likelihood function
analysis would improve the significance by less than a percent.


\section{\label{ConclusionSection} Conclusions and outlook}

In this paper we presented a next-to-leading order study of the
irreducible di-photon background, including the corrections to the
gluon fusion subprocess $gg \rightarrow \gamma \gamma$.  The NLO gluon
fusion is the largest of the higher order contributions not included
in previous studies~\cite{TwoPhotonBkgd1,DIPHOX}.  The scale
dependence of the gluon-fusion contribution at NLO is roughly the same
as at LO if the renormalization and factorization scales are varied
independently, but is significantly reduced if they are varied in
unison.  Moreover, the NLO corrections to the gluon-fusion component,
as a contribution to the total irreducible background, were found to
be modest, suggesting that this calculation is under adequate
theoretical control.  Indeed, the NLO $K$ factor for the $gg \to
\gamma\gamma$ subprocess is only about 65\% of the NLO $K$ factor for
Higgs production.  Experimental studies using the latter $K$ factor to
estimate the former one~\cite{CMS} have therefore been a bit too
conservative.

Using the improved calculation we investigated the statistical
significance of Higgs production as a function of the photon isolation
cuts.  We found that the significance depends only weakly on the
isolation cuts.  Although we found a slight enhancement with more
severe cuts, we noted that as isolation becomes tight, instrumental
noise, soft hadrons and overlapping events can render the cuts experimentally
unworkable.  Moreover, the perturbative predictions become subject to
large corrections and are unreliable.  A better procedure is to
include a veto on jets near the photon candidates.  This can suppress
the background without suffering from the drawbacks of tight photon
isolation.  As an example, we computed the extra suppression due to a
jet veto when the effects of fragmentation can be neglected. We found
that although the background is suppressed, the statistical
significance is hardly altered.  More generally one would need to also
include a jet veto on the fragmentation contribution, but only a weak
dependence may be anticipated.  The most robust improvement we found
in the statistical significance of the Higgs signal, albeit still
modest, was obtained using the rapidity difference $y^*$ distribution
of the decay photons.  It would be interesting to explore other
variables characterizing the distribution of hadronic energy in the
events; a strategy which optimizes the use of this information without
cutting out large numbers of events may be possible.

These results would need to be investigated further in a more
realistic simulation than just the parton-level one we have done here.
In particular, the effects of instrumental noise and overlapping
events must be included~\cite{Wielers}.  One would also need to
include a detailed study of the reducible $\pi^0$ background
contributions~\cite{Tisserand,Wielers,PiBkgd}.  Once a more realistic
study is set up, the entire range up to $m_H < 140$ GeV would need to
be investigated, instead of just the single choice of $m_H = 118$ GeV
used in section~\ref{AngularSubsection}.  When the LHC di-photon data
becomes available, the information provided by the sideband regions
will also be crucial.

As described in the introduction, there are a number of
$\Ord(\alpha_s^2)$ contributions that still have not been included.
These corrections are all expected to be smaller than the
$\Ord(\alpha_s^3)$ gluon fusion contribution incorporated into the
computation presented here.  Nevertheless, for completeness as well as
to confirm that there are no surprises, it would be useful to evaluate
all remaining $\Ord(\alpha_s^2)$ contributions. It would also be
useful to incorporate a resummation of the large logarithms which
appear at the kinematic edges of angular
distributions~\cite{CataniWebber} and for small cone
sizes~\cite{GV,CataniPhotons}.

We are hopeful that further studies will lead to a better
understanding of the di-photon background, and to an increased 
sensitivity for the Higgs search at the LHC.


\begin{acknowledgments}
We thank Thomas Binoth for providing us with a copy of {\tt DIPHOX}.
We also thank Stefano Catani, Fabiola Gianotti, Joey Huston, David
Kosower, Andy Parker, Chris Seez, Zolt\'an Tr\'ocs\'anyi, Monika Wielers, and
C.-P.~Yuan for helpful comments.  L.D. is grateful to DAMTP,
Cambridge, for hospitality while this paper was being completed.

\end{acknowledgments}


\begin{thebibliography}{99}

\bibitem{Higgs}
P.W.~Higgs,
Phys.\ Lett.\  {\bf 12}, 132 (1964),
Phys.\ Rev.\  {\bf 145}, 1156 (1966);\\
F.~Englert and R.~Brout,
Phys.\ Rev.\ Lett.\  {\bf 13}, 321 (1964);\\
G.S.~Guralnik, C.R.~Hagen and T.W.~Kibble,
Phys.\ Rev.\ Lett.\  {\bf 13}, 585 (1964).

\bibitem{HiggsRadCorr}
G.~Degrassi,
arXiv:hep-ph/0102137; \\
J.~Erler,
arXiv:hep-ph/0102143;\\
D. Abbaneo {\it et al.} [ALEPH, DELPHI, L3 and OPAL Collaborations, 
LEP Electroweak Working Group, and SLD Heavy Flavor and Electroweak Groups],
arXiv:hep-ex/0112021.

\bibitem{SusyHiggs}
M.~Carena, H.E.~Haber, S.~Heinemeyer, W.~Hollik, C.E.~Wagner and G.~Weiglein,
Nucl.\ Phys.\ B {\bf 580}, 29 (2000)
[arXiv:hep-ph/0001002]; \\
J.R.~Espinosa and R.~Zhang,
Nucl.\ Phys.\ B {\bf 586}, 3 (2000)
[arXiv:hep-ph/0003246]; \\
A.~Brignole, G.~Degrassi, P.~Slavich and F.~Zwirner,
Nucl.\ Phys.\ B {\bf 631}, 195 (2002)
[arXiv:hep-ph/0112177].

\bibitem{LEP2Limit}
R.~Barate {\it et al.}  [ALEPH Collaboration],
Phys.\ Lett.\ B {\bf 495}, 1 (2000)
[arXiv:hep-ex/0011045];\\
P.~Abreu {\it et al.}  [DELPHI Collaboration],
Phys.\ Lett.\ B {\bf 499}, 23 (2001)
[arXiv:hep-ex/0102036];\\
M.~Acciarri {\it et al.}  [L3 Collaboration],
Phys.\ Lett.\ B {\bf 508}, 225 (2001)
[arXiv:hep-ex/0012019].
G.~Abbiendi {\it et al.}  [OPAL Collaboration],
Phys.\ Lett.\ B {\bf 499}, 38 (2001)
[arXiv:hep-ex/0101014];\\
LEP Higgs Working Group for Higgs boson searches,
Proceedings Intl. Europhysics Conference on High Energy Physics,
Budapest, Hungary, July 2001,
arXiv:hep-ex/0107029.

\bibitem{LEP2MSSMLimit}
LEP Higgs Working Group for Higgs boson searches,
Proceedings Intl. Europhysics Conference on High Energy Physics,
Budapest, Hungary, July 2001,
arXiv:hep-ex/0107030.

\bibitem{RunIIExpectations}
M.~Carena {\it et al.},
arXiv:hep-ph/0010338.

\bibitem{HggVertex}
J.R.~Ellis, M.K.~Gaillard and D.V.~Nanopoulos,
Nucl.\ Phys.\ B {\bf 106}, 292 (1976);\\
M.A.~Shifman, A.I.~Vainshtein, M.B.~Voloshin and V.I.~Zakharov,
Sov.\ J.\ Nucl.\ Phys.\  {\bf 30}, 711 (1979)
[Yad.\ Fiz.\  {\bf 30}, 1368 (1979)].

\bibitem{Higgsgammagamma}
J.F.~Gunion, P.~Kalyniak, M.~Soldate and P.~Galison,
Phys.\ Rev.\ D {\bf 34}, 101 (1986); \\
J.F.~Gunion, G.L.~Kane and J.~Wudka,
Nucl.\ Phys.\ B {\bf 299}, 231 (1988).

\bibitem{HBkgdgammagamma}
R.K.~Ellis, I.~Hinchliffe, M.~Soldate and J.J.~van der Bij,
Nucl.\ Phys.\ B {\bf 297}, 221 (1988).

\bibitem{ATLAS}
ATLAS collaboration,
``ATLAS detector and physics performance, technical design report,'' 
vol. 2, report CERN/LHCC 99-15, ATLAS-TDR-15.

\bibitem{CMS}
CMS collaboration,
``CMS: The electromagnetic calorimeter, technical design report,''
report CERN/LHCC 97-33, CMS-TDR-4.

\bibitem{Tisserand}
V.~Tisserand, 
``The Higgs to two photon decay in the ATLAS detector,''
talk given at the VI International Conference on Calorimetry in
High-Energy Physics, Frascati (Italy), June, 1996, LAL 96-92; Ph.D. thesis,
LAL 97-01, February, 1997.

\bibitem{Wielers}
M.~Wielers, ``Isolation of photons,'' report ATL-PHYS-2002-004.

\bibitem{RZ}
D.~Rainwater and D.~Zeppenfeld,
Phys.\ Rev.\ D {\bf 60}, 113004 (1999)
[Erratum-ibid.\ D {\bf 61}, 099901 (1999)]
[arXiv:hep-ph/9906218];\\
N.~Kauer, T.~Plehn, D.~Rainwater and D.~Zeppenfeld,
Phys.\ Lett.\ B {\bf 503}, 113 (2001)
[arXiv:hep-ph/0012351].

\bibitem{RZH}
D.~Rainwater, D.~Zeppenfeld and K.~Hagiwara,
Phys.\ Rev.\ D {\bf 59}, 014037 (1999)
[arXiv:hep-ph/9808468];\\
T.~Plehn, D.~Rainwater and D.~Zeppenfeld,
Phys.\ Rev.\ D {\bf 61}, 093005 (2000)
[arXiv:hep-ph/9911385].

\bibitem{TwoPhotonBkgd1}
E.L.~Berger, E.~Braaten and R.D.~Field,
Nucl.\ Phys.\ B {\bf 239}, 52 (1984); \\
P.~Aurenche, A.~Douiri, R.~Baier, M.~Fontannaz and D.~Schiff,
Z.\ Phys.\ C {\bf 29}, 459 (1985); \\
B.~Bailey, J.F.~Owens and J.~Ohnemus,
Phys.\ Rev.\ D {\bf 46}, 2018 (1992); \\
B.~Bailey and J.F.~Owens,
Phys.\ Rev.\ D {\bf 47}, 2735 (1993); \\
B.~Bailey and D.~Graudenz,
Phys.\ Rev.\ D {\bf 49}, 1486 (1994)
[arXiv:hep-ph/9307368]; \\
C.~Balazs, E.L.~Berger, S.~Mrenna and C.-P.~Yuan,
Phys.\ Rev.\ D {\bf 57}, 6934 (1998)
[arXiv:hep-ph/9712471]; \\
C.~Balazs and C.-P.~Yuan,
Phys.\ Rev.\ D {\bf 59}, 114007 (1999)
[Erratum-ibid.\ D {\bf 63}, 059902 (1999)]
[arXiv:hep-ph/9810319];\\
T.~Binoth, J.P.~Guillet, E.~Pilon and M.~Werlen,
Phys.\ Rev.\ D {\bf 63}, 114016 (2001)
[arXiv:hep-ph/0012191]; \\
T.~Binoth,
arXiv:hep-ph/0005194.

\bibitem{DIPHOX}
T.~Binoth, J.P.~Guillet, E.~Pilon and M.~Werlen,
Eur.\ Phys.\ J.\ C {\bf 16}, 311 (2000)
[arXiv:hep-ph/9911340]; 

\bibitem{ADW}
L.~Ametller, E.~Gava, N.~Paver and D.~Treleani,
Phys.\ Rev.\ D {\bf 32}, 1699 (1985); \\
%
D.A.~Dicus and S.S.D.~Willenbrock,
Phys.\ Rev.\ D {\bf 37}, 1801 (1988).

\bibitem{GGGamGam}
Z.~Bern, A.~De~Freitas and L.J.~Dixon,
JHEP {\bf 0109}, 037 (2001) [arXiv:hep-ph/0109078].

\bibitem{TwoloopIntegrals}
V.A.~Smirnov,
Phys.\ Lett.\  {\bf B460}, 397 (1999)
[arXiv:hep-ph/9905323];\\
%
V.A.~Smirnov and O.L.~Veretin,
Nucl.\ Phys.\  {\bf B566}, 469 (2000)
[arXiv:hep-ph/9907385];\\
%
J.B.~Tausk,
Phys.\ Lett.\  {\bf B469}, 225 (1999)
[arXiv:hep-ph/9909506];\\
%
C.~Anastasiou, E.W.N.~Glover and C.~Oleari,
Nucl.\ Phys.\  {\bf B565}, 445 (2000) [arXiv:hep-ph/9907523];
Nucl.\ Phys.\  {\bf B575}, 416 (2000)
[arXiv:hep-ph/9912251];\\
%
C.~Anastasiou, T.~Gehrmann, C.~Oleari, E.~Remiddi and J.B.~Tausk,
Nucl.\ Phys.\  {\bf B580}, 577 (2000)
[arXiv:hep-ph/0003261];\\
%
T.~Gehrmann and E.~Remiddi,
Nucl.\ Phys.\  {\bf B580}, 485 (2000)
[arXiv:hep-ph/9912329].

\bibitem{FiveGluon}
Z.~Bern, L.~Dixon and D.A.~Kosower,
Phys.\ Rev.\ Lett.\  {\bf 70}, 2677 (1993)
[arXiv:hep-ph/9302280].

\bibitem{GGGamGamGa}
D.~de~Florian and Z.~Kunszt,
Phys.\ Lett.\ B {\bf 460}, 184 (1999)
[arXiv:hep-ph/9905283].
%

\bibitem{GGGamGamGb}
C.~Balazs, P.~Nadolsky, C.~Schmidt and C.-P.~Yuan,
Phys.\ Lett.\ B {\bf 489}, 157 (2000)
[arXiv:hep-ph/9905551].

\bibitem{CataniSeymour}
S.~Catani and M.H.~Seymour,
Phys.\ Lett.\ B {\bf 378}, 287 (1996)
[arXiv:hep-ph/9602277];
Nucl.\ Phys.\ B {\bf 485}, 291 (1997)
[Erratum-ibid.\ B {\bf 510}, 503 (1997)]
[arXiv:hep-ph/9605323].

\bibitem{PiBkgd}
T.~Binoth, J.P.~Guillet, E.~Pilon and M.~Werlen,
arXiv:hep-ph/0203064.

\bibitem{Frixione}
S.~Frixione,
Phys.\ Lett.\ B {\bf 429}, 369 (1998)
[arXiv:hep-ph/9801442].

\bibitem{AdFS}
K.L.~Adamson, D.~de~Florian and A.~Signer,
Phys.\ Rev.\ D {\bf 65}, 094041 (2002)
[arXiv:hep-ph/0202132].

\bibitem{QQGamGam}
C.~Anastasiou, E.W.N.~Glover and M.E.~Tejeda-Yeomans,
arXiv:hep-ph/0201274.

\bibitem{NLOHKfactor}
M.~Kramer, E.~Laenen and M.~Spira,
Nucl.\ Phys.\ B {\bf 511}, 523 (1998)[arXiv:hep-ph/9611272].

\bibitem{NLOHiggs}
A.~Djouadi, M.~Spira and P.M.~Zerwas,
Phys.\ Lett.\ B {\bf 264}, 440 (1991);\\
S.~Dawson,
Nucl.\ Phys.\ B {\bf 359}, 283 (1991);\\
M.~Spira, A.~Djouadi, D.~Graudenz and P.M.~Zerwas,
Nucl.\ Phys.\ B {\bf 453}, 17 (1995)[arXiv:hep-ph/9504378].

\bibitem{HKNNLO}
R.V.~Harlander and W.B.~Kilgore,
Phys.\ Rev.\ Lett.\  {\bf 88}, 201801 (2002)
[arXiv:hep-ph/0201206].

\bibitem{dFGKRSVN}
D.~de~Florian, M.~Grazzini and Z.~Kunszt,
Phys.\ Rev.\ Lett.\  {\bf 82}, 5209 (1999)
[arXiv:hep-ph/9902483];\\
%
V.~Ravindran, J.~Smith and W.L.~Van Neerven,
arXiv:hep-ph/0201114;\\
%
C.J.~Glosser,
arXiv:hep-ph/0201054.

\bibitem{HDECAY}
A.~Djouadi, J.~Kalinowski and M.~Spira,
Comput.\ Phys.\ Commun.\  {\bf 108}, 56 (1998)
[arXiv:hep-ph/9704448].

\bibitem{MPReview}
M.L.~Mangano and S.J.~Parke,
Phys.\ Rept.\ {\bf 200}, 301 (1991).

\bibitem{VEGAS}
G.P.~Lepage,
J.\ Comput.\ Phys.\  {\bf 27}, 192 (1978).

\bibitem{MRST99}
A.D.~Martin, R.G.~Roberts, W.J.~Stirling and R.S.~Thorne,
Eur.\ Phys.\ J.\ C {\bf 14}, 133 (2000)
[arXiv:hep-ph/9907231].

\bibitem{NLOFrag}
L.~Bourhis, M.~Fontannaz and J.P.~Guillet,
Eur.\ Phys.\ J.\ C {\bf 2}, 529 (1998)
[arXiv:hep-ph/9704447].

\bibitem{CTEQ}
H.L.~Lai {\it et al.}  [CTEQ Collaboration],
Eur.\ Phys.\ J.\ C {\bf 12}, 375 (2000)
[arXiv:hep-ph/9903282].

\bibitem{MRST2001}
A.D.~Martin, R.G.~Roberts, W.J.~Stirling and R.S.~Thorne,
Eur.\ Phys.\ J.\ C {\bf 23}, 73 (2002)
[arXiv:hep-ph/0110215].

\bibitem{CTEQ6M}
J.~Pumplin, D.R.~Stump, J.~Huston, H.L.~Lai, 
P.~Nadolsky and W.K.~Tung,
arXiv:hep-ph/0201195.

\bibitem{CdFG}
S.~Catani, D.~de Florian and M.~Grazzini,
JHEP {\bf 0105}, 025 (2001)
[arXiv:hep-ph/0102227].

\bibitem{gHgamInterf}
Z.~Bern, L.~Dixon and C.~Schmidt, in preparation.

\bibitem{CataniPhotons}
S.~Catani, M.~Fontannaz, J.P.~Guillet and E.~Pilon,
JHEP {\bf 0205}, 028 (2002)
[arXiv:hep-ph/0204023].

\bibitem{GV}
L.E.~Gordon and W.~Vogelsang,
Phys.\ Rev.\ D {\bf 50}, 1901 (1994).

\bibitem{CMS2}
CMS collaboration, Technical Proposal, report CERN/LHCC 94-38.

\bibitem{DD}
M.~Dittmar and H.K.~Dreiner,
Phys.\ Rev.\ D {\bf 55}, 167 (1997)
[arXiv:hep-ph/9608317].

\bibitem{HanZhangCdFG}
T.~Han and R.J.~Zhang,
Phys.\ Rev.\ Lett.\  {\bf 82}, 25 (1999)
[arXiv:hep-ph/9807424]; \\
T.~Han, A.S.~Turcot and R.J.~Zhang,
Phys.\ Rev.\ D {\bf 59}, 093001 (1999)
[arXiv:hep-ph/9812275]; \\
S.~Catani, D.~de Florian and M.~Grazzini,
JHEP {\bf 0201}, 015 (2002)
[arXiv:hep-ph/0111164].

\bibitem{Abdullin}
S.~Abdullin, M.~Dubinin, V.~Ilyin, D.~Kovalenko, V.~Savrin and 
N.~Stepanov,
Phys.\ Lett.\ B {\bf 431}, 410 (1998)
[arXiv:hep-ph/9805341].

\bibitem{BY}
C.~Balazs and C.-P.~Yuan,
Phys.\ Lett.\ B {\bf 478}, 192 (2000)
[arXiv:hep-ph/0001103].

\bibitem{IsolatedPhotonIR}
E.L.~Berger and J.W.~Qiu,
Phys.\ Rev.\ D {\bf 44}, 2002 (1991);\\
%
S.~Catani, M.~Fontannaz and E.~Pilon,
Phys.\ Rev.\ D {\bf 58}, 094025 (1998)
[arXiv:hep-ph/9803475].

\bibitem{CataniWebber}
S.~Catani and B.R.~Webber,
JHEP {\bf 9710}, 005 (1997)
[arXiv:hep-ph/9710333].

\end{thebibliography}


\end{document}